\documentclass[aps,11pt]{revtex4}
\usepackage{amssymb}
\usepackage{amsmath}
\usepackage{graphicx}
\usepackage{epsfig}
\usepackage{amsfonts}
\usepackage{dcolumn}
\usepackage{bm}

\begin{document}

\title{ On the relative quantum entanglement \\
with respect to tensor product structure}
\author{X. F. Liu $^{1}$, and C. P. Sun $^{2}$}
\affiliation{ $^{1}$ Department of Mathematics,Peking University,
Beijing 100871, China}
 \affiliation{ $^{2}$ Institute of Theoretical Physics, The Chinese
Academy of Science, Beijing, 100080, China}

\begin{abstract}
Mathematical foundation of the novel concept of quantum tensor
product by Zanardi et al is rigorously established. The concept of
relative quantum entanglement is naturally introduced and its
meaning is made clear both mathematically and physically. For a
finite or an infinite dimensional vector space $W$ the so called
tensor product partition (TPP) is introduced on $End(W)$, the set
of endmorphisms of $W$, and a natural correspondence is
constructed between the set of TPP's of $End(W)$ and the set of
tensor product structures (TPS's) of $W$. As a byproduct, it is
shown that an arbitrarily given wave function belonging to an
n-dimensional Hilbert space, $n$ being not a prime number, can be
interpreted  as a separable state with respect to some man-made
TPS, and thus a quantum entangled state of a many-body system with
respect to the ``God-given" TPS can be regarded as a quantum state
without entanglement in some sense. The concept of standard set of
observables is also introduced to probe the underlying structure
of the object TPP and to establish its connection with practical
physical measurement.

\end{abstract}

\maketitle

\section{Introduction}

Quantum entanglement is a fundamental concept of quantum mechanics
and plays a central role in quantum information processing
\cite{qi}. It has also motivated many investigations in
mathematical physics\cite{q-math,MBR}. What is less obvious is the
fact that quantum entanglement is not an intrinsically defined
concept. For example, the state of a bi-particle system described
by the separable wave function with respect to the two position
coordinates is generally an entangled one with respect to the
center of mass and relative coordinates. It seems that this point
has been ignored for a long time by physicists. But recently
Zanardi et.al have brought this problem to our attention in the
context of quantum information. They explicitly point out that
whether a state is an entangled one or not depends on the tensor
product structure (TPS) of the state space \cite{z1,z2} and they
argue that quantum system can be partitioned into the so called
virtual subsystems according to a man-made TPS selected by a set
of observables operationally relevant in the sense of interactions
and measurements. Accordingly, quantum entanglement is observable
induced and hence relative. As a matter of fact, we have in some
sense considered the relativity of quantum entanglement in the
adiabatic separation of the fast and slow variables of a composite
system by means of the Born-Oppenheimer approximation
\cite{sun1,sun2}.

In this paper, to characterize the above mentioned man-made TPS and the
related quantum entanglement we develop a general algebraic approach from the
view point of observable algebra. Along this line, we manage to show that a
quantum state can justifiably be called entangled or unentangled with respect
to a particular partition of the observable algebra. And among others, we can
show that in an $n$-dimensional Hilbert space $W$ where the dimension $n$ is
not a prime, an arbitrarily given quantum state $|s\rangle\in W$ is a
separable one with respect to some man-made TPS of $W=\mathbf{V}_{k^{\prime}%
}\otimes^{\prime}\mathbf{V}_{l^{\prime}}$ where $\mathbf{V}_{k^{\prime}%
},\mathbf{V}_{l^{\prime}}$ are respectively $k^{\prime}$ and $l^{\prime}$
dimensional subspaces of $W$ . Particularly, an entangled state in $W$ with
respect to the natural (or ``a priori God-given" ) TPS $\mathbf{W}=\mathbf{V}%
_{k}\otimes\mathbf{V}_{l}$ with $n=kl$ , can always be decoded as
a separable state with respect to some artificially introduced
TPS. But this is only one side of the coin. It is in fact equally
true that when the dimension of $W$ is not a prime an arbitrarily
given quantum state $|s\rangle\in W$ is an entangled one with
respect to some man-made TPS. In short, we have made it
mathematically clear that it is impossible to make a clear cut
between entanglement and unentanglement as expected. To emphasize
the physical aspect of the TPS of state space we also show how it
is related to the so called complete set of observables.

It should be pointed out that the main idea of this paper
originates from the remarkable observations implied in the
interesting paper by Zanardi et al\cite{z1,z2}. \ But in this
paper we prove the uniqueness theorem concerning the TPS related
to a particular partition of the observable algebra while only the
existence theorem is proved in the original paper. Thanks to the
uniqueness theorem we can characterize entanglement from the view
point of partition of observable algebra. Moreover, the whole
theory is here developed within strict mathematical framework and
thus some vague points have been clarified. Especially, we have
weakened the conditions for the partition of observable algebra
and developed a method which is free of the restriction from
dimension. Indeed, most of the results obtained in this paper are
valid in the infinite dimensional case as well as in the finite
dimensional case. As we do not wish to restrict ourselves to the
finite dimensional case from the very beginning some proofs become
inevitably more complicated. But our effort is rewarded: we
finally clarify the arguments about the entanglement of identical
particle system. It is found that different definitions of
entanglement for distinguishable-particle tacitly presuppose
different TPS's corresponding to the measurement of different
observables, and from our approach a\ separability criterion for
two-identical particle system can be correctly given without any
contradiction\cite{you,long}.

The rest parts of this paper is organized as follows. \ In section
II we present some basic knowledge of module theory and
multilinear algebra for later use and fix the notation. The
materials are standard and can be found in any relevant text books
(for example see \cite{K}). The reader who is familiar with these
topics can safely skip this section. In section III, we introduce
and investigate the concept of tensor product partition (TPP) of
the set of linear operators $End(W)$ on a finite or an infinite
dimensional vector space $W$, which turns out to have a close
connection with\ the TPS of $W$. In section IV we introduce the
concept of complete set of operators to probe the underlying
structure of the TPP of $End(W).$ With these preparations we study
the interrelationship between the TPP of $End(W)$ and the TPS of
$W$ in section V. Some major propositions are proved in this
section concerning the correspondence between these two objects. \
For the application in quantum mechanics, we take into accounts
the inner product structure of $W$ in section VI and discuss the
inner product compatible TPS after introducing a natural
compatibility condition. In section VII, we explore the
relationship among the TPP, the TPS and the product vector set,
and the relativity of entanglement then becomes clear.\ Finally,
three examples are analyzed in section VIII as an illustration of
the theory and some concluding remarks are made in section IX.

\section{Preliminaries}

In this paper all the algebras and vector spaces dealt with are over the
complex number field and of countable dimension. Moreover, we will not
consider the topology of vector space at all. So infinite summation does not
mean any limit process. Rather, its meaning will be specified in the context.

First let us review an elementary part of the module theory. Let $A$ be an
associative algebra, $V$ an $A$ module. $V$ is called a irreducible module if
it has no nontrivial submodule. If $V$ can be decomposed into a direct sum of
irreducible modules, then it is called decomposable.

\textbf{Theorem 2.1} \textit{Let }$V$\textit{\ be an }$A$\textit{\ module,
then (1) }$V$\textit{\ is decomposable if and only if every submodule of }%
$V$\textit{\ is decomposable; (2) }$V$\textit{\ is decomposable only if for
any submodule }$V_{1}$\textit{\ of }$V$\textit{\ there is a complementary
submodule }$V_{2}$\textit{: }$V=V_{1}\oplus V_{2}.$

\textit{Remark} If every submodule of $V$\ is finitely generated, then the
converse of (2) in the above theorem is true. That is, if for any submodule of
$V$\ there exists a complementary submodule, then $V$\ is decomposable.

\textbf{Theorem 2.2} \textit{Let }$V$\textit{\ be a decomposable }%
$A$\textit{\ module, }$V=%
{\textstyle\sum_{i}}
\oplus V_{i}$\textit{\ the decomposition of }$V$\textit{\ into a direct sum of
irreducible submodules. If }$U$\textit{\ is a irreducible submodule of }%
$V,$\textit{\ then }$U$\textit{\ is isomorphic to some }$V_{i}.$

Let $V$ and $W$ be $A$ modules. A linear map from $V$ to $W$ is
called\textit{\ a module homomorphism} if it commutes with the action of $A.$
An injective and surjective module homomorphism is called a module
isomorphism. The following result concerning module homomorphism is well known.

\textbf{Theorem 2.3} \textit{Let }$V$\textit{\ and }$W$\textit{\ be
irreducible }$A$\textit{\ modules, }$f$\textit{\ a homomorphism from }%
$V$\textit{\ to }$W$\textit{. Then }$f$\textit{\ is either a zero map or an
isomorphism.}

According to this theorem, given two irreducible modules, to prove that they
are isomorphic we only need to show that there exists a non zero homomorphism
between them. This fact will be used in the next section.

\textbf{Definition 2.1 }An irreducible $A$ module $V$ will be called a normal
module if the following condition is satisfied: every homomorphism $f$ from
$V$ to itself is equal to the identity map up to a scalar multiple.

\textit{Remark}\textbf{\ }A finite dimensional irreducible $A$\ module is
necessarily a normal module by Schur's Lemma. But a normal module is not
necessarily finite dimensional. So normality in the sense of Definition 2.1
does not characterize finite dimensionality completely. In this paper we will
assume the normality, instead of the finite dimensionality, of certain
modules, and hence our discussion is applicable to some interesting infinite
dimensional cases.

Next let us recall some basic knowledge concerning the concept of tensor product.

\textbf{Definition 2.2} \textit{Let }$V_{1},V_{2}$\textit{\ and }%
$W$\textit{\ be vector spaces, }$f$\textit{\ a bilinear map from }%
$(V_{1},V_{2})$\textit{\ to }$W.$\textit{\ If for any vector space }%
$U$\textit{\ and any bilinear map }$g$\textit{\ from }$(V_{1},V_{2}%
)$\textit{\ to }$U$\textit{\ there exists a unique linear map }$h$%
\textit{\ from }$W$\textit{\ to }$U$\textit{\ such that }$g=h\circ
f,$\textit{\ then }$(V_{1},V_{2},f)$\textit{\ is called a TPS of }%
$W$\textit{\ and }$W$\textit{\ is called a tensor product of }$V_{1}%
$\textit{\ and }$V_{2}$\textit{\ with respect to }$f,$\textit{\ or simply a
tensor product of }$V_{1}$\textit{\ and }$V_{2}$\textit{\ if no confusion will
arise.}

Conventionally, $f$ is denoted by the symbol $\otimes$ and $W$ is
written as $W=V_{1}\otimes V_{2}.$ Here is a major fact about
tensor product: If $\left\{  x_{i}\right\}  $ and $\left\{
y_{j}\right\}  $ are two bases of $V_{1}$ and $V_{2}$
respectively, then $W=V_{1}\otimes V_{2}$ if and only if $\left\{
x_{i}\otimes y_{j}\right\}  $ is a basis of $W.$ \ \ Especially in
the finite dimensional case, $(V_{1},V_{2},f)$ is a TPS of $W$ if
and only if the image of $f$ spans $W$ and $\dim W=\dim
V_{1}\cdot\dim V_{2}.$

\textbf{Definition 2.3} \textit{Let }$(V_{1},V_{2},\otimes)$\textit{\ be a TPS
of }$W.$\textit{\ A vector }$w\in W$\textit{\ is called decomposable if it is
of the form }$x\otimes y$\textit{\ where }$x\in V_{1}$\textit{\ and }$y\in
V_{2}.$\textit{\ }

Obviously, any vector $w\in W$ can be written as a sum of decomposable
vectors. We call such a sum an expression of $w$ in terms of decomposable
vectors or just an expression of $w$ for short. Notice that expressions of $w
$ may not be unique. The length of an expression of $w$ is defined to be the
number of nonzero decomposable vectors it contains and the rank of $w$ is
defined to be the length of the shortest expressions of $w.$ Then by
definition the rank of a decomposable vector is $1.$ The following result
about the rank of a vector is useful.

\textbf{Proposition 2.1} \textit{The expression }$w=%
{\textstyle\sum_{i,j}}
u_{i}\otimes v_{j}$\textit{\ is a shortest one if and only if }$\left\{
u_{i}\right\}  $\textit{\ and }$\left\{  v_{j}\right\}  $\textit{\ are
linearly independent in }$V_{1}$\textit{\ and }$V_{2}$\textit{\ respectively}.

Finally we consider $End(W),$ the set of endmorphisms of $W.$ Take $a\in
End(V_{1})$ and $b\in End(V_{2}).$ Then we can define a bilinear map $g$ from
$(V_{1},V_{2})$ to $W$ such that $g(u,v)=au\otimes bv$ $\forall u\in
V_{1},v\in V_{2}.$ So there is a unique endmorphism $h$ of $W$ such that
$h(u\otimes v)=au\otimes bv.$ By convention such an $h$ will be denoted by
$a\otimes b$ from now on. Thus we have $(a\otimes b)(u\otimes v)=au\otimes bv.
$ Denote by $S$ the linear subspace of $End(W)$ spanned by the endmorphisms of
the form $a\otimes b.$ Then $(End(V_{1}),End(V_{2}),\otimes)$ is a TPS of $S.$
Here, $\otimes$ stands for the bilinear map satisfying $\otimes(a,b)=a\otimes
b$ as the symbol itself suggests. In the finite dimensional case we have
$S=End(W),$ so $End(W)=End(V_{1})\otimes End(V_{2}).$ But when $W$ is of
infinite dimension, this is no longer true. We will investigate this problem
in more detail in the next sections.

\section{Tensor Product Partition}

In this section and the next one, we introduce the concept of TPP for the
endmorphisms of the finite or infinite dimensional vector space $W$. In this
section, $A$ always stands for $End(W).$ The concept of TPP is at the core of
this paper. It turns out to be useful in understanding relativity of quantum entanglement.\

\textbf{Definition 3.1} \ \ (a) \textit{For }$a_{i}\in A,$\textit{\ the
summation }$%
{\textstyle\sum_{i}}
a_{i}$\textit{\ is called well defined if }$\left(
{\textstyle\sum_{i}}
a_{i}\right)  w$\textit{\ is a well defined vector of }$W.$ (b)
\textit{A subset }$B$\textit{\ of }$A$\textit{\ is called an
extended subalgebra if it is a subalgebra in the usual sense and
is closed under well defined summation.}

Let $A_{1}$ and $A_{2}$ be two extended subalgebras of $A$ such
that $\left[ A_{1},A_{2}\right]  =0,$ namely $\left[  a,b\right]
=0$ for $a\in A_{1}$ and $b\in A_{2}.$ We denote by $A_{1}\vee
A_{2}$ the associative algebra generated by $A_{1}$ and $A_{2}.$
By definition $A_{1}\vee A_{2}\subseteq A$ and an element $c$ of
$A$ belongs to $A_{1}\vee A_{2}$ if and only if $c$ is of the
well defined summation form $%
{\textstyle\sum_{i}}
a_{i}b_{i}$ where $a_{i}\in A_{1}$ and $b_{i}\in A_{2}.$ Notice that the sum $%
{\textstyle\sum_{i}}
a_{i}b_{i}$ may contain infinitely many terms but $\left(
{\textstyle\sum_{i}}
a_{i}b_{i}\right)  w$ contains only finite terms for each $w\in W.$

\textbf{Definition 3.2} \textit{The ordered pair }$\left(  A_{1},A_{2}\right)
$\textit{\ is called a pre-tensor-product partition of }$A$\textit{\ if the
following two conditions are satisfied: (1) }$\left[  A_{1},A_{2}\right]
=0$\textit{\ and }$A=A_{1}\vee A_{2};$\textit{\ (2) }$W$\textit{\ is a
decomposable }$A_{1}$\textit{\ and }$A_{2}$\textit{\ modules respectively.}

For arbitrary extended subalgebras $A_{1}$ and $A_{2}$ of $A,$ $W$
becomes $A_{1}$ and
$A_{2}$ modules in the natural way. We find that when $\left(  A_{1}%
,A_{2}\right)  $ is a pre-tensor-product partition of $A$ the modules enjoy a
very nice property.

\textbf{Lemma 3.1} \textit{If }$\left(  A_{1},A_{2}\right)  $\textit{\ is a
pre-tensor-product partition of }$A,$\textit{\ then all irreducible }$A_{1}%
$(\textit{\ }$A_{2}$)\textit{\ submodules of }$W$\textit{\ are isomorphic .}

\textit{Proof}. Since $W$ is a decomposable $A_{1}$ module, we have the
decomposition of $W$ into a direct sum of irreducible $A_{1}$ modules:
\[
W=\sum_{i}\oplus M_{i}.
\]
It follows that each irreducible $A_{1}$ submodule is isomorphic
to some $M_{i}.$ We need to show that all $M_{i}$'s are isomorphic
to one another. For different indices $i,j,$ we choose $c\in A$
such that $cM_{i}\subseteq M_{j}$ and $c|_{M_{i}},$ the
restriction of $c$ to $M_{i},$
is nonzero. As $A=A_{1}\vee A_{2}$ we can write $c=%
{\textstyle\sum_{k}}
$ $a_{k}b_{k}$ where $a_{k}\in A_{1}$ and $b_{k}\in A_{2}.$ Denote by $p_{l}$
the projection onto $M_{l}.$ Obviously, $p_{l}$ is an $A_{1}$ module
homomorphism and we have $%
{\textstyle\sum_{l}}
p_{l}=1,$ where $1$ stands for the identity map. Now $c$ can be rewritten as
\[
c=\sum_{k,l}a_{k}p_{l}b_{k}.
\]
Notice that $p_{l}b_{k}M_{i}\subseteq M_{l}$ and $a_{k}p_{l}b_{k}%
M_{i}\subseteq M_{l}$ for each $l.$ So it follows from $cM_{i}\subseteq M_{j}$
that
\[
c|_{M_{i}}=\sum_{k}a_{k}p_{j}b_{k}.
\]
But $c|_{M_{i}}\neq0$, thus there exists a $k$ such that $p_{j}b_{k}\neq0.$
Finally $b_{k}\in A_{2}$ implies that $b_{k}$ is an $A_{1}$ module
homomorphism. Therefore $p_{j}b_{k}$ is a nonzero $A_{1}$ module homomorphism
from $M_{i}$ to $M_{j}$ , and $M_{i}$ and $M_{j}$ are isomorphic according to
Theorem 2.3. This proves the lemma for $A_{1}.$ The parallel result for
$A_{2}$ can be proved in the same way.

\ \ \ \ \ \ \ \ \ \ \ \ \ \ \ \ \ \ \ \ \ \ \ \ \ \ \ \ \ \ \ \ \ \ \ \ \ \ \ \ \ \ \ \ \ \ \ \ \ \ \ \ \ \ \ \ \ \ \ $\square
$

For a pre-tensor-product partition $\left(  A_{1},A_{2}\right)  ,$ by
definition we have the decompositions%
\[
W=\sum_{i}\oplus M_{i}=\sum_{j}\oplus N_{j},
\]
where $M_{i}$ and $N_{j}$ are irreducible $A_{1}$ and $A_{2}$ modules
respectively. According to Lemma 2.1, all $M_{i}$'s and all $N_{j}$'s are
isomorphic. This allows us to denote them by $M$ and $N$ respectively. For
convenience, $M,N$ will be called \textit{characteristic modules}, and
$\left\{  M_{i}\right\}  ,\left\{  N_{j}\right\}  $ irreducible component
sets, of the partition $\left(  A_{1},A_{2}\right)  .$

\textbf{Definition 3.3} \textit{Let }$\left(  A_{1},A_{2}\right)
$\textit{\ be a pre-tensor-product partition of }$A.$\textit{\ It is called a
TPP if its characteristic modules are normal modules.}

\textit{Remark} In the finite dimensional case the concepts of
pre-tensor-product partition and TPP are equivalent. But in the
infinite dimensional cases a pre-tensor-product partition of\ $A$\
may not be a TPP of\ $A.$This is because Schur's Lemma may be
false in the infinite dimensional case.

\textbf{Lemma 3.2} \textit{If }$\left(  A_{1},A_{2}\right)  $\textit{\ is a
TPP, then }$A_{1}|_{M_{i}}=End(M_{i})$\textit{\ and }$A_{2}|_{N_{j}}%
=End(N_{j}),$\textit{\ where }$\left\{  M_{i}\right\}  $\textit{\
( }$\left\{ N_{j}\right\}  $\textit{\ ) is an irreducible
component set of }$A_{1}$ ($A_{2}$)$.$

\textit{Proof.} The notation is the same as in the proof of Lemma 3.1, unless
explicitly pointed out. We use the contradiction method. If $A_{1}|_{M_{i}%
}\subsetneqq End(M_{i}),$ then there exists an element $c_{i}\in End(M_{i})$
which can not be written as $a|_{M_{i}}$ with $a\in A_{1}.$ Take $c\in A$ such
that $c|_{M_{i}}=c_{i}.$We can write
\[
c=%
{\textstyle\sum_{k}}
a_{k}b_{k}=\sum_{k,l}a_{k}p_{l}b_{k}.
\]
It then follows that
\[
c|_{M_{i}}=\sum_{k}a_{k}p_{i}b_{k}.
\]
On the other hand, $p_{i}b_{k}|_{M_{i}}$ is an $A_{1}$ module homomorphism
from $M_{i}$ to $M_{i},$ so there is a constant $\alpha_{k}$ such that
$p_{i}b_{k}|_{M_{i}}=\alpha_{k}\bullet1$ because $M_{i}$ is a normal module by
assumption. Thus we can write%
\[
c|_{M_{i}}=\left(  \sum_{k}\alpha_{k}a_{k}\right)  |_{M_{i}}.
\]
Now we define $a=\sum_{k}\alpha_{k}a_{k}.$ We claim that $a\in A_{1}.$ To
prove this point, it suffices to show that $a$ is well defined. In other
words, we only need to show that $aw$ contains only finite terms for each
$w\in W.$ Note that $a|_{M_{i}}=c|_{M_{i}}$ is well defined and $W=%
{\textstyle\sum_{i}}
\oplus M_{i}.$ The well definedness of $a$ then follows directly from the fact
that all $M_{i}$'s are isomorphic $A_{1}$ modules. As $a|_{M_{i}}=c_{i}$ we
are led to a contradiction. This proves $A_{1}|_{M_{i}}=End(M_{i}).$ The other
conclusion can be proved in the same way.\ \

\ \ \ \ \ \ \ \ \ \ \ \ \ \ \ \ \ \ \

\ \ \ \ \ \ \ \ \ \ \ \ \ \ \ \ \ \ \ \ \ \ \ \ \ \ \ \ \ \ \ \ \ \ \ \ \ \ \ \ \ \ \ \ \ \ \ \ \ \ \ \ \ \ \ \ \ \ \ \ \ \ $\square
$

Let $\left(  A_{1},A_{2}\right)  $ be a TPP, $\left\{  M_{i}\right\}  $ the
irreducible component set for $A_{1}.$ According to\textit{\ Lemma 3.1}, for
each $i$ we can choose an ordered basis $\left\{  x_{ji}|j=1,2,\cdots\right\}
$ of $M_{i}$ such that $A_{1}$ has the same matrix representation with respect
to these bases. In other words, for each $a\in A_{1}$ there exists a complex
number set $\left\{  a_{kl}|k,l=1,2,\cdots\right\}  ,$ which is independent of
the index $i,$ such that
\[
ax_{ji}=\sum_{k}x_{ki}a_{kj},\forall i.
\]
Obviously, $%
{\textstyle\bigcup_{i}}
\left\{  x_{ji}|j=1,2,\cdots\right\}  $ is a basis of $W.$ We call such an
ordered basis \textit{a synchronic basis} with respect to the irreducible
component set $\left\{  M_{i}\right\}  .$Actually, $%
{\textstyle\bigcup_{i}}
\left\{  x_{ji}|j=1,2,\cdots\right\}  $ is a synchronic basis with respect to
the irreducible component set $\left\{  M_{i}\right\}  $ if and only if the
linear map $f_{i}$ from $M_{1}$ to $M_{i}$ that sends $x_{j1} $ to $x_{ji}$ is
a module isomorphism.

Now for each $j$ define $N_{j}$ to be the vector space spanned by $\left\{
x_{ji}|i=1,2,\cdots\right\}  .$ Then we have the following result.

\textbf{Lemma 3.3} \textit{(1) }$\left\{  N_{j}\right\}  $\textit{\ is the
irreducible component set for }$A_{2};$\textit{\ (2) }$%
{\textstyle\bigcup_{j}}
\left\{  x_{ji}|i=1,2,\cdots\right\}  $\textit{\ is a synchronic basis with
respect to }$\left\{  N_{j}\right\}  .$

\textit{Proof.} (1) Take a set $\left\{  \lambda_{j}|j=1,2,\cdots\right\}  $
consisting of distinct complex numbers and define $r_{i}\in End(M_{i})$ such
that $r_{i}x_{ji}=\lambda_{j}x_{ji}.$ Then according to \textit{Lemma 3.2,}
there is $r\in A_{1}$ such that $r|_{M_{i}}=r_{i}.$ As $%
{\textstyle\bigcup_{i}}
\left\{  x_{ji}|j=1,2,\cdots\right\}  $ is a synchronic basis we
have $rx_{ji}=\lambda_{j}x_{ji},\forall i.$ It follows that
$N_{j}$ is none other than the eigenspace of $r$ corresponding to
the eigenvalue $\lambda_{j}.$ But $\left[  r,A_{2}\right] =0 $ by
definition, so $N_{j}$ is stable under the action of $A_{2},$ that
is, $N_{j}$ is an $A_{2}$ module. Clearly we have
\[
W=\sum_{i}\oplus M_{i}=\sum_{j}\oplus N_{j}.
\]
Now it remains to show that $N_{j}$ is an irreducible module.

If $N_{j}$ is not irreducible, then by Theorem 1.1 it can be decomposed into a
direct sum of at least two irreducible modules:%
\[
N_{j}=\sum_{k}\oplus N_{j_{k}}.
\]
Consider $c\in A=End(W)$ such that $cN_{j_{1}}\subseteq N_{j_{2}}$ and
$c|_{N_{j_{1}}}\neq0.$ We claim that $c\notin A_{1}\vee A_{2}.$ This
contradicts the condition $A_{1}\vee A_{2}=End(W).$ Hence, it is sufficient to
prove the claim.

Denote by $p_{j}$ the projection to $N_{j}.$ We first prove that for each
$a\in A_{1}$ and $z\in$ $N_{j}$ there exists a complex number $\alpha$ such
that $p_{j}az=\alpha z.$ In fact, for $z\in$ $N_{j}$ we can write%
\[
z=\sum_{i}\alpha_{i}x_{ji},\alpha_{i}\in%
\Bbb{C} .
\]
Because $%
{\textstyle\bigcup_{i}}
\left\{  x_{ji}|j=1,2,\cdots\right\}  $ is a synchronic basis with respect to
$\left\{  M_{i}\right\}  ,$ the action of $a$ is of the form
\[
ax_{ji}=\sum_{k}x_{ki}a_{kj},\forall i,
\]
where $a_{kj}$ is independent of $i.$Thus%
\[
az=\sum_{i}\sum_{k}\alpha_{i}x_{ki}a_{kj},
\]
and%
\[
p_{j}az=a_{jj}\sum_{i}\alpha_{i}x_{ji}=a_{jj}z.
\]
Now we prove $c\notin A_{1}\vee A_{2}.$ If $c\in A_{1}\vee A_{2},$ then there
exist $a_{k}\in A_{1},b_{k}\in A_{2}$ such that
\[
c=\sum_{k}a_{k}b_{k}.
\]
By assumption $cN_{j_{1}}\subseteq N_{j_{2}}\subsetneqq N_{j},$ so for $z\in
N_{j_{1}}$ we have
\[
cz=p_{j}cz=\sum_{k}p_{j}a_{k}b_{k}z.
\]
Define $z_{k}=b_{k}z.$ Obviously, $z_{k}\in N_{j_{1}}\subsetneqq
N_{j}.$ Then by the above argument for each $k$ there exists a
complex number $\alpha_{k}$ such that
$p_{j}a_{k}z_{k}=\alpha_{k}z_{k}.$ It now follows that
\[
cz=\sum_{k}p_{j}a_{k}z_{k}=\sum_{k}\alpha_{k}z_{k}\in N_{j_{1}}.
\]
But $cz\in N_{j_{2}}$ by definition, so $cz=0,\forall z\in N_{j_{1}}.$ This is
a contradiction. The proof for the first part of the lemma is thus completed.

(2) As $N_{j}$ is an $A_{2}$ module, for $b\in A_{2}$ most generally the
action on the basis element $x_{ji}$ can be written as%
\[
bx_{ji}=\sum_{k}x_{jk}b_{ki}^{j},
\]
where $b_{ki}^{j}\in%
\Bbb{C}
$ depends on the index $j.$ Suppose that $%
{\textstyle\bigcup_{j}}
\left\{  x_{ji}|i=1,2,\cdots\right\}  $ is not a synchronic basis
with respect to $\left\{  N_{j}\right\}  .$ Then there exist an
element $b\in A_{2}$ and indices $j_{1},j_{2},k,i$ such that
$b_{ki}^{j_{1}}\neq b_{ki}^{j_{2}}.$ Since
$A_{1}|_{M_{i}}=End(M_{i}),$ there exists an element $a\in A_{1}$
satisfying $ax_{j_{1}i}=x_{j_{2}i}.$ It then follows that
$ax_{j_{1}k}=x_{j_{2}k},\forall
k$ because $%
{\textstyle\bigcup_{i}}
\left\{  x_{ji}|j=1,2,\cdots\right\}  $ is a synchronic basis with respect to
$\left\{  M_{i}\right\}  .$ Now consider $ab$ and $ba.$ We have%
\begin{align*}
abx_{j_{1}i}  &  =\sum_{k}ax_{j_{1}k}b_{ki}^{j_{1}}=\sum_{k}x_{j_{2}k}%
b_{ki}^{j_{1}},\\
bax_{j_{1}i}  &  =bx_{j_{2}i}=\sum_{k}x_{j_{2}k}b_{ki}^{j_{2}}.
\end{align*}
So we come to the conclusion that $abx_{j_{1}i}\neq bax_{j_{1}i}.$ But this is
impossible since $\left[  A_{1},A_{2}\right]  =0$ by definition. The proof is
thus completed.

\bigskip
\ \ \ \ \ \ \ \ \ \ \ \ \ \ \ \ \ \ \ \ \ \ \ \ \ \ \ \ \ \ \ \ \ \ \ \ \ \ \ \ \ \ \ \ \ \ \ \ \ \ \ \ \ \ \ $\square
$

\textbf{Corollary 3.1}. \textit{Let }$\left(  A_{1},A_{2}\right)
$\textit{\ be a TPP of }$End(W),$\textit{\ then there exist an
irreducible component set }$\left\{  M_{i}\right\}  $\textit{\ for
}$A_{1},$\textit{\ an
irreducible component set }$\left\{  N_{j}\right\}  $\textit{\ for }$A_{2}%
,$\textit{\ and a basis }$\left\{  x_{ji}\right\}  $\textit{\ such that }$%
{\textstyle\bigcup_{i}}
\left\{  x_{ji}|j=1,2,\cdots\right\}  $\textit{\ is a synchronic basis with
respect to }$\left\{  M_{i}\right\}  $\textit{\ and }$%
{\textstyle\bigcup_{j}}
\left\{  x_{ji}|i=1,2,\cdots\right\}  $\textit{\ a synchronic basis with
respect to }$\left\{  N_{j}\right\}  .$

This corollary follows directly from the above lemma. A basis $\left\{
x_{ji}\right\}  $ with the property specified in the corollary will be
called\textit{\ a standard basis} associated with the irreducible component
sets $\left\{  M_{i}\right\}  $ and $\left\{  N_{j}\right\}  .$

\textbf{Corollary 3.2. }\textit{Let }$\left(  A_{1},A_{2}\right)
$\textit{\ be a TPP of }$End(W),$\textit{\ }$\left\{  \lambda_{j}\right\}
$\textit{\ and }$\left\{  \mu_{i}\right\}  $\textit{\ two sets of distinct
complex numbers. Then there exist }$r\in A_{1}$\textit{, }$t\in A_{2}%
$\textit{, and a decompositions of }$W$\textit{\ into direct sum
of vector
spaces:}%
\[
W=\sum_{i}\oplus M_{i}=\sum_{j}\oplus N_{j}%
\]
\textit{such that }%
\[
r|_{N_{j}}=\lambda_{j}\cdot1,\text{ }t|_{M_{i}}=\mu_{i}\cdot1
\]
\textit{and the standard module decomposition }%
\[
M_{i}=\sum_{j}\oplus M_{i}^{\lambda_{j}},\text{ }N_{j}=\sum_{i}\oplus
N_{j}^{\mu_{i}}%
\]
\textit{where the summations range over all }$\lambda_{j}$\textit{'s and all
}$\mu_{i}$\textit{'s respectively, and }$M_{i}^{\lambda_{j}}$\textit{\ and
}$N_{j}^{\mu_{i\text{ }}}$\textit{\ are one dimensional. Here }$M_{i}%
^{\lambda_{j}}$\textit{\ stands for the eigenspace of }$r$\textit{\ in }%
$M_{i}$\textit{\ corresponding to the eigenvalue
}$\lambda_{j}$\textit{\ and
}$N_{j}^{\mu_{i\text{ }}}$\textit{\ the eigenspace of }$t$\textit{\ in }%
$N_{j}$\textit{\ corresponding to the eigenvalue }$\mu_{i}.$

\textit{Proof.} According to Corollary 3.1, there exist an
irreducible component set $\left\{  M_{i}\right\}  $ for $A_{1},$
an irreducible component set $\left\{  N_{j}\right\}  $ for
$A_{2},$ and a basis $\left\{
x_{ji}\right\}  $ such that $%
{\textstyle\bigcup_{i}}
\left\{  x_{ji}|j=1,2,\cdots\right\}  $ is a synchronic basis with respect to
$\left\{  M_{i}\right\}  $ and $%
{\textstyle\bigcup_{j}}
\left\{  x_{ji}|i=1,2,\cdots\right\}  $ a synchronic basis with respect to
$\left\{  N_{j}\right\}  .$ Define $r,t\in A$ such that%
\[
rx_{ji}=\lambda_{j}x_{ji},\text{ }tx_{ji}=\mu_{i}x_{ji}.
\]
It is easy to verify that $\left\{  M_{i}\right\}  ,\left\{  N_{j}\right\}
,r,t$ meet the requirement of Corollary 3.2.

\section{Standard Complete Set of Operators}

In the last section, some fine properties for TPP have been proved to prepare
for the introduction of the TPS. In this section we proceed along to probe the
underlying structure of TPP, aiming at describing the TPS of a finite or
infinite dimensional vector space $W$ in a constructive way. We will show how
the TPP of $End\left(  W\right)  $ is determined by particular sets of
operators, the so called standard complete sets of operators, contained in
$End\left(  W\right)  $.

\textbf{Definition 4.1} I\textit{f }$r,t\in A$\textit{\ satisfy the conditions
specified in Corollary 3.2, then }$\left(  r,t\right)  $\textit{\ is called a
standard complete set of operators of }$A$\textit{, and }$\left\{
M_{i}\right\}  $\textit{, }$\left\{  N_{j}\right\}  $\textit{\ are called
characteristic sets of }$r$\textit{\ and }$t$\textit{\ respectively.}

\textbf{Definition 4.2} \textit{If }$\left(  r,t\right)  $\textit{\ is a
standard complete set of operators of }$A$\textit{\ and }$\left(  A_{1}%
,A_{2}\right)  $\textit{\ is a TPP of }$A$\textit{\ such that }$r\in A_{1}%
$\textit{\ and }$t\in A_{2},$\textit{\ then }$\left(  A_{1},A_{2}\right)
$\textit{\ is called a TPP containing }$\left(  r,t\right)  .$

It is obvious that, if $\left(  r,t\right)  $ is a standard
complete set of operators, then necessarily $\left[  r,t\right]
=0$ according to the above definitions.

We have seen that a TPP contains standard complete sets of operators. Now it
is natural to ask how to determine a TPP from a standard complete set of
operators. The remaining part of this section is devoted to this problem.

\textbf{Proposition 4.1 }If $\left(  r,t\right)  $ is a standard complete set
of operators of $A,$ then there exists a TPP $\left(  A_{1},A_{2}\right)  $
containing $\left(  r,t\right)  .$

\textit{Proof}. By the definition of complete set of operators, there are two
sets of distinct complex numbers $\left\{  \lambda_{j}\right\}  $ and
$\left\{  \mu_{i}\right\}  $ and two decompositions of $W$ into direct sum of
subspaces
\[
W=\sum_{i}\oplus M_{i}=\sum_{j}\oplus N_{j}%
\]
such that
\[
M_{i}=\sum_{j}\oplus%
\Bbb{C}
x_{ji},\text{ }N_{j}=\sum_{i}\oplus%
\Bbb{C}
x_{ji}%
\]
where $x_{ji}$ is the common eigenvector of $r$ and $t$:
\[
rx_{ji}=\lambda_{j}x_{ji},\text{ }tx_{ji}=\mu_{i}x_{ji}.
\]
It then follows that we can define two extended subalgebras
$A_{1},A_{2}\subseteq A$
such that the following two conditions are satisfied: (1) $A_{1}|_{M_{i}%
}=End(M_{i})$ and $A_{2}|_{N_{j}}=End(N_{j});$ (2) $%
{\textstyle\bigcup_{i}}
\left\{  x_{ji}|j=1,2,\cdots\right\}  $ becomes a synchronic basis with
respect to $\left\{  M_{i}\right\}  $ and $%
{\textstyle\bigcup_{j}}
\left\{  x_{ji}|i=1,2,\cdots\right\}  $ a synchronic basis with respect to
$\left\{  N_{j}\right\}  .$We claim that $\left(  A_{1},A_{2}\right)  $ is a
TPP and $r\in A_{1}$, $t\in A_{2}.$ The claim is almost immediate from the
definition. In fact, the first condition guarantees that $W$ are decomposable
$A_{1}$ and $A_{2}$ modules, and $M_{i}$, $N_{j}$ are respectively irreducible
normal $A_{1}$ and $A_{2}$ modules, while the second condition leads to the
commutation relation $\left[  A_{1},A_{2}\right]  =0.$ The fact that $r\in
A_{1}$ and $t\in A_{2}$ is also a direct consequence of the two conditions.
Now it remains to show that $A_{1}\vee A_{2}=A.$ This point is proved as follows.

Let $a\in A$ be an arbitrary element. It suffices to prove that $a\in
A_{1}\vee A_{2}.$Define $a_{k,l},b_{k,l}\in A$ such that%
\[
a_{k,l}x_{ji}=\delta_{kj}x_{li},\text{ }b_{k,l}x_{ji}=\delta_{ki}x_{jl}.
\]
It is readily verified that $\left\{  a_{k,l}\right\}  $ and $\left\{
b_{k,l}\right\}  $ are bases of $A_{1}$ and $A_{2}$ respectively. As $\left\{
x_{ji}\right\}  $ is a basis of $W$, $a$ is determined by its action on each
$x_{ji}.$ Generally we can write
\[
ax_{ji}=\sum_{k,l}x_{kl}a_{kl,ji},a_{kl,ji}\in%
\Bbb{C} .
\]
Notice that for each $x_{ji},$ there are only finite nonzero coefficients
$a_{kl,ji}.$ One can now easily convince oneself that the expression $%
{\textstyle\sum_{j,i}}
{\textstyle\sum_{k,l}}
a_{kl,ji}a_{j,k}b_{i,l}$ is a well defined summation and equal to the given
$a.$ Obviously this is an element of $A_{1}\vee A_{2}. $ The proposition is
thus proved.

\bigskip

\ \ \ \ \ \ \ \ \ \ \ \ \ \ \ \ \ \ \ \ \ \ \ \ \ \ \ \ \ \ \ $\ \ \ \ \ \ \ \ \ \ \ \ \ \ \ \ \ \ \ \ \ \ \ \ \ \square
$

Notice that Proposition 4.1 solves the problem of existence of TPP containing
a given complete set of operators. To probe the problem of uniqueness in some
sense, we need to make some more preparation. The next proposition is also
interesting in its own right.

\textbf{Proposition 4.2} \textit{If }$\left(  A_{1},A_{2}\right)
$\textit{\ is a TPP of }$A,$\textit{\ then }$A_{1}^{\prime}=A_{2},$%
\textit{\ }$A_{2}^{\prime}=A_{1}.$\textit{\ Here }$A_{i}^{\prime}$%
\textit{\ }$\left(  i=1,2\right)  $\textit{\ stands for the commutator of
}$A_{i}$\textit{\ in }$A $\textit{: }$A_{i}^{\prime}=\left\{  a\in A|\left[
a,A_{i}\right]  =0\right\}  .$

\textit{Proof. }Let $\left\{  M_{i}\right\}  $ and $\left\{
N_{j}\right\}  $ be irreducible component sets of $\left(
A_{1},A_{2}\right)  ,$ $\left\{ x_{ji}\right\}  $ a synchronic
basis associated with $\left\{  M_{i}\right\}  $ and $\left\{
N_{j}\right\}  .$ Define $a_{k,l},b_{k,l}\in A$ in the same way as
in the proof of \textit{Proposition 4.1}. Now we prove the
proposition in three steps as follows.

(1) If $%
{\textstyle\sum_{l}}
{\textstyle\sum_{l_{k}}}
a\left(  l_{k},l\right)  b_{l_{k},l}=0,$ then $%
{\textstyle\sum_{l_{k}}}
a\left(  l_{k},l\right)  b_{l_{k},l}=0$ for each $l,$ where $a\left(
l_{k},l\right)  \in A_{1}.$ In fact, we have%
\[
\left(
{\textstyle\sum_{l_{k}}}
a\left(  l_{k},l\right)  b_{l_{k},l}\right)  W\subseteq M_{l},\forall l,
\]
so the conclusion directly follows from the decomposition $W=\sum_{i}\oplus
M_{i}.$

(2) If $%
{\textstyle\sum_{k}}
a_{k}b_{k,l}=0,$ then $a_{k}=0$ for each $k,$ where $a_{k}\in A_{1}.$ If, on
the contrary, there is some $a_{i}\neq0,$ then $a_{i} $ can be written as%
\[
a_{i}=\sum_{k,l}\alpha_{k,l}a_{k,l},
\]
where there is at least a nonzero coefficient. Suppose that $\alpha_{m,n}%
\neq0.$ Then it is readily check that $\left(
{\textstyle\sum_{k}}
a_{k}b_{k,l}\right)  x_{m,i}\neq0.$ This contradicts the condition $%
{\textstyle\sum_{k}}
a_{k}b_{k,l}=0.$ The statement is thus proved.

(3) $A_{2}=A_{1}^{\prime},$ $A_{1}=A_{2}^{\prime}.$ By definition
$A_{2}\subseteq A_{1}^{\prime}.$ So to prove $A_{2}=A_{1}^{\prime},$ we only
need to show that $A_{1}^{\prime}\subseteq A_{2}.$ Let $a\in A_{1}^{\prime
}\subseteq A.$ As $A_{1}\vee A_{2}=A$ we can express $a$ in the form%
\[
a=%
{\textstyle\sum_{l}}
{\textstyle\sum_{l_{k}}}
a\left(  l_{k},l\right)  b_{l_{k},l},
\]
where $a\left(  l_{k},l\right)  \in A_{1}.$ To prove that $a\in A_{2}$ it is
sufficient to show that $a\left(  l_{k},l\right)  $ is equal to the identity
map up to a scalar multiple. In fact, if this is not the case, then there
exist an $a\left(  l_{k},l\right)  $ and some $c\in A_{1}$ such that $\left[
c,a\left(  l_{k},l\right)  \right]  \neq0$ because $A_{1}|_{M_{i}}=End(M_{i})$
for every $i.$ On the other hand, we have
\[
0=\left[  c,a\right]  =%
{\textstyle\sum_{l}}
{\textstyle\sum_{l_{k}}}
\left[  c,a\left(  l_{k},l\right)  \right]  b_{l_{k},l}.
\]
It then follows from (1) and (2) that $\left[  c,a\left(  l_{k},l\right)
\right]  =0,$ for every $l$ and $l_{k}.$ This contradiction proves that
$A_{2}=A_{1}^{\prime}$. Similarly, we can prove that $A_{1}=A_{2}^{\prime}.$

\ \ \ \ \ \ \ \ \ \ \ \ \ \ \ \ \ \ \ \ \ \ \ \ \ \ \ \ \ \ \ \ \ \ \ \ \ \ \ \ \ \ \ \ \ \ \ \ \ \ \ \ \ \ \ \ $\square
$

\textbf{Corollary 4.1 } If $\left(  A_{1},A_{2}\right)  $ is a TPP of $A,$
then $A_{1}\cap A_{1}^{\prime}=A_{2}\cap A_{2}^{\prime}=%
\Bbb{C} 1.$

\textit{Proof.} First we notice that it is a direct consequence of Proposition
3.2 that $A_{1}\cap A_{1}^{\prime}=A_{2}\cap A_{2}^{\prime}$. Let $\left\{
M_{i}\right\}  $ and $\left\{  N_{j}\right\}  $ be irreducible component sets
for $A_{1}$ and $A_{2}$ respectively, and $\left\{  x_{ji}\right\}  $ a
synchronic basis associated with them. By definition%
\[
M_{i}=\sum_{j}\oplus%
\Bbb{C}
x_{ji},\text{ }N_{j}=\sum_{i}\oplus%
\Bbb{C} x_{ji}.
\]
As $A_{1}|_{M_{i}}=End(M_{i})$ and $A_{2}|_{N_{j}}=End(N_{j})$ it is clear
that $%
\Bbb{C} 1\subseteq A_{1}\cap A_{2}=A_{1}\cap A_{1}^{\prime}.$ For
the same reason, if $a\in A_{1}\cap A_{1}^{\prime}$ $=A_{2}\cap
A_{2}^{\prime},$ then there exist constant sets $\left\{
\alpha_{i}\right\}  ,\left\{  \beta_{j}\right\}
\subseteq%
\Bbb{C} $ such that $a|_{M_{i}}=\alpha_{i}\cdot1,$
$a|_{N_{j}}=\beta_{j}\cdot1.$ It
then follows that all these constants are identical. Thus $a\in%
\Bbb{C}
1,$ that is, $A_{1}\cap A_{1}^{\prime}$ $=A_{2}\cap A_{2}^{\prime}\subseteq%
\Bbb{C} 1.$ This completes the proof.

\ \ \ \ \ \ \ \ \ \ \ \ \ \ \ \ \ \ \ \ \ \ \ \ \ \ \ \ \ \ \ \ \ \ \ \ \ \ \ \ \ \ \ \ \ \ \ \ \ \ \ \ \ $\square
$

\textit{Remark} In the finite dimensional case, from this corollary we
conclude that if $\left(  A_{1},A_{2}\right)  $\ is a TPP, then both $A_{1}%
$\ and $A_{2}$\ are the\ so called factors.

\textbf{Lemma 4.1} \textit{Let }$W$\textit{\ be a vector space, }%
$q$\textit{\ a linear transformation of }$W.$\textit{\ If }$q$\textit{\ is
diagonalizable and all of its eigenvalues are distinct, then }$q$\textit{\ is
diagonalizable in any }$q$\textit{\ invariant subspace of }$W.$

Before proving this lemma we remark that the conclusion is obvious if $W$ is
finite dimensional, but if this is not the case the lemma seems to need a
proof. Certainly, we present the lemma and its proof here not to claim the
originality. Rather, we do so just for completeness.

\textit{Proof.} Let $\left\{  \lambda_{j}\right\}  $ be the set of eigenvalues
of $q.$ Then we have the decomposition%
\[
W=\sum_{j}%
\Bbb{C} x_{j},
\]
where $x_{j}$ is an eigenvector of $q$ corresponding to the eigenvalue
$\lambda_{j}$: $qx_{j}=\lambda_{j}x_{j}.$ Suppose that $W_{1}\subseteq W$ is a
$q$ invariant subspace, namely, $qW_{1}\subseteq W_{1}.$ For an arbitrary
$y\in W_{1},$ we can write%
\[
y=\sum_{j}\alpha_{j}\left(  y\right)  x_{j},\text{ }\alpha_{j}\left(
y\right)  \in%
\Bbb{C} .
\]
If there exists $y\in W_{1}$ such that $\alpha_{j}\left(  y\right)  \neq0,$
then we call $\lambda_{j}$ an eigenvalue related to the subspace $W_{1}.$ We
claim that
\[
W_{1}=\sum_{\lambda_{j}}\oplus W^{\lambda_{j}}%
\]
where $W^{\lambda_{j}}=%
\Bbb{C} x_{j}$ and the summation ranges over all the eigenvalues
that are related to $W_{1}.$ Clearly we have
\[
W_{1}\subseteq\sum_{\lambda_{j}}\oplus W^{\lambda_{j}}.
\]
So to prove the claim it is sufficient to show that for each eigenvalue
$\lambda_{j}$ that is related to $W_{1}$ we have $x_{j}\in W_{1}.$ In fact, if
$\lambda_{j_{0}}$ is related to $W_{1},$ then there exists $y\in W_{1}$ such
that
\[
y=\sum_{j\in I}\alpha_{j}\left(  y\right)  x_{j}%
\]
where $I$ is a finite set containing $j_{0}$ and $\alpha_{j}\left(  y\right)
\neq0,$ $\forall j\in I.$ Suppose that $I$ contains $n$ elements. Then we have
the following system of linear equations:%
\[
q^{i}y=\sum_{j\in I}\alpha_{j}\left(  y\right)  \lambda_{j}^{i}x_{j},\text{
}i=1,2,\cdots,n.
\]
As $qW_{1}\subseteq W_{1},$ we have $q^{i}y\in W_{1},$ $\forall i\in I.$ On
the other hand, the determinant of the coefficient matrix is nonzero since all
the $\lambda_{j}$'s are distinct. Therefore, we have $x_{j}\in W_{1},$
$\forall j\in I,$ especially, $x_{j_{0}}\in W_{1}.$ This proves the claim, and
hence the lemma.

\bigskip
\ \ \ \ \ \ \ \ \ \ \ \ \ \ \ \ \ \ \ \ \ \ \ \ \ \ \ \ \ \ \ \ \ \ \ \ \ \ \ \ \ \ \ \ \ \ \ \ \ \ \ \ \ \ \ \ $\square
$

\textbf{Lemma 4.2} \textit{Let }$\left(  r,t\right)  $\textit{\ be a standard
complete set of operators of }$A\left(  =End(W)\right)  $\textit{, }$\left\{
M_{i}\right\}  $\textit{, }$\left\{  N_{j}\right\}  $\textit{\ the
characteristic sets of }$r$\textit{\ and }$t$\textit{\ respectively. If
}$\left(  A_{1},A_{2}\right)  $\textit{\ is a TPP containing }$\left(
r,t\right)  ,$\textit{\ then }$\left\{  M_{i}\right\}  $\textit{, }$\left\{
N_{j}\right\}  $\textit{\ are irreducible component sets for }$A_{1}%
$\textit{\ and }$A_{2}$\textit{\ respectively.}

\textit{Proof.} By definition we have%
\[
W=\sum_{i}\oplus M_{i}=\sum_{j}\oplus N_{j},
\]
and%
\[
r|_{N_{j}}=\lambda_{j}\cdot1,\text{ }t|_{M_{i}}=\mu_{i}\cdot1.
\]
Let us focus on $\left\{  M_{i}\right\}  .$ Notice that $M_{i}$ is none other
than the eigenspace of $t$ corresponding to the eigenvalue $\mu_{i}.$ As $t\in
A_{2}$, we have $\left[  t,A_{1}\right]  =0.$ It then follows that $A_{1}%
M_{i}\subseteq M_{i},$ that is, $M_{i}$ is an $A_{1}$ module. We observe that
proving that $\left\{  M_{i}\right\}  $ is a irreducible component set for
$A_{1}$ boils down to proving that $M_{i}$ is irreducible as $A_{1}$ module.
The proof is as follows.

Suppose that $M_{i}$ is not irreducible. Then $M_{i}$ can be decomposed into a
direct sum of nonzero irreducible modules:%
\[
M_{i}=\sum_{k}\oplus M_{i_{k}}.
\]
By Lemma 4.1 all $M_{i_{k}}$'s are isomorphic. On the other hand, according to
Lemma 4.1, $r$ is diagonalizable in each $M_{i_{k}}.$ Note that $r\in$
$A_{1}.$ Thus $r$ has the same eigenvalues in different $M_{i_{k}}$'s. But
this is impossible because by definition all the eigenvalues of $r$ in $M_{i}
$ have the multiplicity $1.$ In the same way we can prove that $N_{j}$ is a
irreducible $A_{2}$ module.

\bigskip
\ \ \ \ \ \ \ \ \ \ \ \ \ \ \ \ \ \ \ \ \ \ \ \ \ \ \ \ \ \ \ \ \ \ \ \ \ \ \ \ \ \ \ \ \ \ \ \ \ \ \ \ $\square
$

Now we are prepared to prove the following result concerning the uniqueness of
TPP containing a given complete set of operators.

\textbf{Proposition 4.3} \textit{Let }$\left(  r,t\right)  $\textit{\ be a
standard complete set of operators of }$A\left(  =End(W)\right)  ,$%
\textit{\ }$\left(  A_{1},A_{2}\right)  $\textit{\ and }$\left(  B_{1}%
,B_{2}\right)  $\textit{\ two tensor product partitions containing }$\left(
r,t\right)  .$\textit{\ Then there exists an isomorphism }$\varphi\in
End(W)$\textit{, diagonal with respect to the basis consisting of common
eigenvectors of }$r$\textit{\ and }$t,$\textit{\ such that }$B_{1}%
=\varphi\cdot A_{1}\cdot\varphi^{-1}$\textit{\ and }$B_{2}=\varphi\cdot
A_{2}\cdot\varphi^{-1}.$

\textit{Proof.} Keep the same notation as in the proof of Lemma 4.2. According
to Lemma 4.2, $\left\{  M_{i}\right\}  $ is a irreducible component set for
both $A_{1}$ and $B_{1}.$ Fix an index $i_{0}$ and choose a basis $\left\{
x_{ji_{0}}\right\}  $ of $M_{i_{0}}$ such that $rx_{ji_{0}}=\lambda
_{j}x_{ji_{0}}$. Obviously we can extend this basis to a synchronic basis $%
{\textstyle\bigcup_{i}}
\left\{  x_{ji}|j=1,2,\cdots\right\}  $ with respect to $\left\{
M_{i}\right\}  $ as irreducible component set for $A_{1}$ and a synchronic
basis $%
{\textstyle\bigcup_{i}}
\left\{  y_{ji}|j=1,2,\cdots\right\}  $ with respect to $\left\{
M_{i}\right\}  $ as irreducible component set for $B_{1}$. Since
$r\in A_{1},$ $B_{1},$ we have
\[
rx_{ji}=\lambda_{j}x_{ji},\text{ }ry_{ji}=\lambda_{j}y_{ji},\text{ }\forall i.
\]
It then follows that for each pair of index $\left(  i,j\right)  $ there
exists a complex number $\alpha_{ji}$ such that $y_{ji}=$ $\alpha_{ji}x_{ji}.$
This is because that all the eigenvalues of $r$ in $M_{i}$ are of multiplicity
$1.$ Now define $\varphi\in End(W)$ such that $\varphi x_{ji}=y_{ji}.$ Then
$\varphi$ is an isomorphism diagonal with respect to the basis $\left\{
x_{ji}\right\}  .$ It is clear that $B_{1}=\varphi\cdot A_{1}\cdot\varphi
^{-1}.$ Indeed, this relation follows directly from the fact that
$A_{1}|_{M_{i}}=B_{1}|_{M_{i}}=End(M_{i}).$ Finally, we consider the set
$\varphi\cdot A_{2}\cdot\varphi^{-1}.$ We have $\left[  B_{1},\varphi\cdot
A_{2}\cdot\varphi^{-1}\right]  =0$, so by Proposition 4.2 $\varphi\cdot
A_{2}\cdot\varphi^{-1}\subseteq B_{2}.$ Similarly, we can prove $\varphi
^{-1}\cdot B_{2}\cdot\varphi\subseteq A_{2}.$ Thus $B_{2}=\varphi\cdot
A_{2}\cdot\varphi^{-1}.$ This completes the proof.

\bigskip
\ \ \ \ \ \ \ \ \ \ \ \ \ \ \ \ \ \ \ \ \ \ \ \ \ \ \ \ \ \ \ \ \ \ \ \ \ \ \ \ \ \ \ \ \ \ \ \ \ \ \ \ \ \ \ \ $\square
$

We have seen that a TPP is determined up to an isomorphism by a
standard complete set of operators contained in it. Now, in the
remaining part of this section, we study how to determine a TPP
completely by some standard complete sets of operators satisfying
certain conditions. For convenience, we first introduce a new
concept as follows. For $p,q\in End(W),$ we denote by $S_{p,q}$
the extended subalgebra of $End(W)$ generated by them. Let $\left(
r,t\right)  ,$ $\left(  r^{\prime},t^{\prime}\right)  $ be
standard complete sets of operators with the characteristic sets
$\left\{  M_{i}\right\} ,\left\{  N_{j}\right\}  $ and $\left\{
M_{i}^{\prime}\right\}  ,\left\{ N_{j}^{\prime}\right\}  $
respectively.

\textbf{Definition 4.3} $\left(  r,t\right)  $ and $\left(  r^{\prime
},t^{\prime}\right)  $ are called complementary if (1) $M_{i}=M_{i}^{\prime} $
and all $M_{i}$'s are isomorphic normal $S_{r,r^{\prime}}$ modules or (2)
$N_{j}=N_{j}^{\prime}$ and all $N_{j}$'s are isomorphic normal $S_{t,t^{\prime
}}$ modules.

\textit{Remark} Both (1) and (2) cannot be satisfied unless $M_{i}$\ and
$N_{j}$\ are both of one dimension.

Next we prove the following results on the construction of TPP.

\textbf{Proposition 4.4} \textit{A TPP contains complementary standard
complete sets of operators.}

\textit{Proof}. Let $\left(  A_{1},A_{2}\right)  $ be a TPP with the
irreducible component sets $\left\{  M_{i}\right\}  ,\left\{  N_{j}\right\}
.$ Take a synchronic basis $\left\{  x_{ji}\right\}  $ associated with
$\left\{  M_{i}\right\}  ,\left\{  N_{j}\right\}  .$ Then there exists a
standard complete set of operators $\left(  r,t\right)  $ with the
characteristic sets $\left\{  M_{i}\right\}  ,\left\{  N_{j}\right\}
:rx_{ji}=\lambda_{j}x_{ji},$ $tx_{ji}=\mu_{i}x_{ji}.$ Now define
$\widetilde{r},\widetilde{t}\in End(W)$ such that:%
\begin{align*}
\widetilde{t} &  =t,\\
\widetilde{r}x_{1i} &  =\lambda_{1}x_{1i},\text{ }\widetilde{r}\left(
x_{ji}+x_{j+1\text{ }i}\right)  =\lambda_{j+1}\left(  x_{ji}+x_{j+1\text{ }%
i}\right)  .
\end{align*}
It is readily check that $\left(  \widetilde{r},\widetilde{t}\right)  $ is a
standard complete set of operators contained in $\left(  A_{1},A_{2}\right)
$. Obviously all $M_{i}$'s are isomorphic $S_{r,\widetilde{r}}$ modules by
definition. Now to prove the proposition it suffices to show that $M_{i}$ is a
normal $S_{r,\widetilde{r}}$ module. Let $f:M_{i}\longrightarrow M_{i}$ be an
$S_{r,\widetilde{r}}$ module homomorphism. Then we have%
\begin{align*}
rf\left(  x_{ji}\right)   &  =\lambda_{j}f\left(  x_{ji}\right)  ,\\
\widetilde{r}f\left(  x_{1i}\right)   &  =\lambda_{1}f\left(  x_{1i}\right)
,\widetilde{r}f\left(  x_{ji}+x_{j+1\text{ }i}\right)  =\lambda_{j+1}f\left(
x_{ji}+x_{j+1\text{ }i}\right)  .
\end{align*}
It follows that there are $\alpha_{j},\beta_{j}\in%
\Bbb{C}
$ such that%
\[
f\left(  x_{ji}\right)  =\alpha_{j}x_{ji},\text{ }f\left(  x_{ji}+x_{j+1\text{
}i}\right)  =\beta_{j+1}\left(  x_{ji}+x_{j+1\text{ }i}\right)  .
\]
We thus conclude that all $\alpha_{j}$'s must be identical, that is,
$f=\alpha\cdot1$ for some $\alpha\in%
\Bbb{C} .$ Hence, $M_{i}$ is a normal $S_{r,\widetilde{r}}$
module.

\ \ \ \ \ \ \ \ \ \ \ \ \ \ \ \ \ \ \ \ \ \ \ \ \ \ \ \ \ \ \ \ \ \ \ \ \ \ \ \ \ \ \ \ \ \ \ \ \ \ \ \ $\square
$

\textbf{Proposition 4.5}\textit{\ If }$\left(  r,t\right)  $\textit{\ and
}$\left(  \widetilde{r},\widetilde{t}\right)  $\textit{\ are complementary
standard complete sets of operators, then there exists a unique TPP }$\left(
A_{1},A_{2}\right)  $\textit{\ such that }$r,\widetilde{r}\in$\textit{\ }%
$A_{1}$\textit{\ and }$t,\widetilde{t}\in A_{2}.$

\textit{Proof.}\textbf{\ }Let us consider the case where $r,\widetilde{r}$
have the same characteristic set $\left\{  M_{i}\right\}  $ and all $M_{i}$'s
are isomorphic normal $S_{r,\widetilde{r}}$ modules. The other case can be
discussed in the same way.

Let $f_{i}:M_{1}\longrightarrow M_{i}$ be an $S_{r,\widetilde{r}}$ module
isomorphism. By the definition of standard complete set of operators, there
exist bases $\left\{  x_{j1}\right\}  ,\left\{  \widetilde{x}_{j1}\right\}  $
of $M_{1}$ and sets $\left\{  \lambda_{j}\right\}  ,\left\{  \widetilde
{\lambda}_{j}\right\}  $ of distinct complex numbers such that
\[
rx_{j1}=\lambda_{j}x_{j1},\text{ }\widetilde{r}\widetilde{x}_{j1}%
=\widetilde{\lambda}_{j}\widetilde{x}_{j1}.
\]
Let
$x_{ji}=f_{i}x_{j1},\widetilde{x}_{ji}=f_{i}\widetilde{x}_{j1}.$
Then according to the proof of Proposition 3.1, there are TPP's
$\left(  A_{1},A_{2}\right)  $ and $\left(
\widetilde{A}_{1},\widetilde
{A}_{2}\right)  $ such that (1) $r\in A_{1},\widetilde{r}\in\widetilde{A}%
_{1},t\in A_{2},\widetilde{t}\in\widetilde{A}_{2};$ (2) $\cup_{i}\left\{
x_{ji}|j=1,2,\cdots\right\}  $ and $\cup_{i}\left\{  \widetilde{x}%
_{ji}|j=1,2,\cdots\right\}  $ are synchronic bases with respect to $\left\{
M_{i}\right\}  $ as irreducible component sets for $A_{1}$ and $\widetilde
{A}_{1}$ respectively. Since $f_{i}$ is an $S_{r,\widetilde{r}}$ module
isomorphism, $\cup_{i}\left\{  \widetilde{x}_{ji}|j=1,2,\cdots\right\}  $ is
also a synchronic basis with respect to $\left\{  M_{i}\right\}  $ as
irreducible component sets for $A_{1}.$ It then follows that $A_{1}%
=\widetilde{A}_{1}$ and hence that $A_{2}=\widetilde{A}_{2}$ as $A_{2}%
=A_{1}^{\prime}$ and $\widetilde{A}_{2}=\widetilde{A}_{1}^{\prime}.$ Thus
$\left(  A_{1},A_{2}\right)  $ is a TPP meeting the requirement. This proves
the existence.

Now let $\left(  B_{1},B_{2}\right)  $ be an arbitrary TPP satisfying the
condition. To prove the uniqueness we only need to show that $\left(
B_{1},B_{2}\right)  =\left(  A_{1},A_{2}\right)  ,$ which is defined above.
According to Lemma 4.2, $\left\{  M_{i}\right\}  $ is an irreducible component
set for $B_{1}.$ Then there exists a synchronic basis $\cup_{i}\left\{
y_{ji}|j=1,2,\cdots\right\}  $ with respect to $\left\{  M_{i}\right\}  $ such
that $y_{j1}=x_{j1,\text{ }}j=1,2,\cdots.$ As $r,\widetilde{r}\in B_{1}$ the
linear map $g_{i}:$ $M_{1}\longrightarrow M_{i}$ that sends $y_{j1}$ to
$y_{ji}$ is an $S_{r,\widetilde{r}}$ module isomorphism. But $M_{1}$ is a
normal $S_{r,\widetilde{r}}$ module, so there exists $\alpha_{i}\in%
\Bbb{C} $ such that $f_{i}^{-1}\cdot g_{i}=\alpha_{i}\cdot1,$ and
we have $y_{ji}=\alpha_{i}x_{ji}.$ It then follows that
$\cup_{i}\left\{ x_{ji}|j=1,2,\cdots\right\}  $ is also a
synchronic basis with respect to $\left\{  M_{i}\right\}  $ as
irreducible component set for $B_{1}.$ Consequently, we have
$A_{1}=B_{1}$ and hence $A_{2}=B_{2}.$ The uniqueness is thus
proved.

\ \ \ \ \ \ \ \ \ \ \ \ \ \ \ \ \ \ \ \ \ \ \ \ \ \ \ \ \ \ \ \ \ \ \ \ \ \ \ \ \ \ \ \ \ \ \ \ \ \ \ \ \ \ \ \ \ \ \ \ \ \ \ \ $\square
$

\section{Tensor Product Structure}

With the above preparation in concepts we are now in a position to focus on
the TPS of a vector space $W,$ one of the mainstay of this paper. In this
section we will establish a correspondence between the set of TPS of $W$ and
the set of TPP of $End(W),$ revealing the close relation between these two
objects. In this section we denote $End(W)$ by $A.$

\textbf{Definition 5.1} \textit{Let }$(W_{1},W_{2},\otimes)$\textit{\ be a TPS
of }$W,$ $\left(  A_{1},A_{2}\right)  $\textit{\ a TPP of }$A\left(
=End(W)\right)  ,$\textit{\ \ where }$W_{1},W_{2}$\textit{\ are subspaces of
}$W.$\textit{\ }$(W_{1},W_{2},\otimes)$\textit{\ is called a TPS associated
with }$\left(  A_{1},A_{2}\right)  $\textit{\ if the following condition is
satisfied:}%
\begin{align*}
a(u\otimes v) &  =\left(  au\right)  \otimes v,\text{ }b(u\otimes
v)=u\otimes\left(  bv\right)  ,\\
\text{ }\forall a &  \in A_{1},b\in A_{2},u\in W_{1},v\in W_{2}.
\end{align*}

According to this definition, if $(W_{1},W_{2},\otimes)$ is a TPS associated
with $\left(  A_{1},A_{2}\right)  ,$ then $W_{1}$ and $W_{2}$ are necessarily
$A_{1}$ and $A_{2}$ modules respectively. Furthermore, we have the following result.

\textbf{Lemma 5.1}\textit{\ if }$(W_{1},W_{2},\otimes)$\textit{\ is a TPS
associated with }$\left(  A_{1},A_{2}\right)  ,$\textit{\ then }$W_{1}%
$\textit{\ and }$W_{2}$\textit{\ are irreducible }$A_{1}$\textit{\ and }%
$A_{2}$\textit{\ modules respectively}.

\textit{Proof}. Suppose, on the contrary, that $W_{1}$ is not
irreducible. Then there exist nonzero $A_{1}$ modules
$W_{1}^{\alpha}$ and $W_{1}^{\beta}$ such that
$W_{1}=W_{1}^{\alpha}\oplus W_{1}^{\beta}.$ Take two nonzero
elements $x_{1}\in W_{1}^{\alpha},$ $x_{2}\in W_{1}^{\beta}$ and
an element $a\in A$ such that $ax_{1}=x_{2}.$ It is clear that
$a\notin A_{1}\vee A_{2}. $ This contradicts the condition that
$A_{1}\vee A_{2}=A.$ That $W_{2} $ is irreducible can be proved in
the same way.

\ \ \ \ \ \ \ \ \ \ \ \ \ \ \ \ \ \ \ \ \ \ \ \ \ \ \ \ \ \ \ \ \ \ \ \ \ \ \ \ \ \ \ \ \ \ \ \ \ \ \ \ \ \ $\square
$

\ \ \ \ \ \ \ \ \ \ \ \ \ \ \ \ \ \ \ \ \ \ \ \ \ \ \ \ \ \ \ \ \ \ \ \ \ \ \ \ \ \ \ \ \ \ \ \ \ \ \ \ \ \ \ \

\textbf{Proposition 5.1} \textit{Let }$(W_{1},W_{2},\otimes)$\textit{\ be a
TPS of }$W,$\textit{\ where }$W_{1},W_{2}$\textit{\ are subspaces of }%
$W.$\textit{\ Define }$A_{1}=End(W_{1})\otimes1\triangleq\left\{
a\otimes1|a\in End(W_{1})\right\}  $\textit{\ and }$A_{2}=1\otimes
End(W_{2})\triangleq\left\{  1\otimes b|b\in End(W_{2})\right\}
.$\textit{\ Then }$\left(  A_{1},A_{2}\right)  $\textit{\ is a TPP of }%
$A$\textit{\ and }$(W_{1},W_{2},\otimes)$\textit{\ is a TPS associated with
}$\left(  A_{1},A_{2}\right)  .$\textit{\ Conversely, if }$\left(  A_{1}%
,A_{2}\right)  $\textit{\ is a TPP of }$A$\textit{\ and }$(W_{1},W_{2}%
,\otimes)$\textit{\ a TPS associated with it, then we have }$A_{1}=\left\{
a\otimes1|a\in End(W_{1})\right\}  $\textit{\ and }$A_{2}=\left\{  1\otimes
b|b\in End(W_{2})\right\}  .$

\textit{Proof.} The proof of the first part is immediate, and we would rather
omit it. For the second part, just notice that if $\left\{  x_{j}\right\}  $
and $\left\{  y_{i}\right\}  $ are respective bases of $W_{1}$ and $W_{2},$
then $\left\{  W_{1}\otimes y_{i}\right\}  $, $\left\{  x_{j}\otimes
W_{2}\right\}  $ are irreducible component sets for $A_{1}$ and $A_{2}$
respectively and $\left\{  x_{j}\otimes y_{i}\right\}  $ is a standard basis
associated with them. The conclusion then follows. Here $W_{1}\otimes
y_{i}=\left\{  u\otimes y_{i}|u\in W_{1}\right\}  $ as the symbol suggests,
and $x_{j}\otimes W_{2}$ is understood similarly.

\ \ \ \ \ \ \ \ \ \ \ \ \ \ \ \ \ \ \ \ \ \ \ \ \ \ \ \ \ \ \ \ \ \ \ \ \ \ \ \ \ \ \ \ \ \ \ \ \ \ \ \ \ \ \ \ $\square
$

This proposition tells us that each TPS of the form $(W_{1},W_{2},\otimes)$
with $W_{1},W_{2}\subseteq W$ is associated with some TPP determined by it.
Naturally we want to ask whether a TPP can determine a TPS associated with it.
The answer is positive.

\textbf{Theorem 5.1} \textit{Each TPP of }$A$\textit{\ determines a TPS
associated with it.}

\textit{Proof.} Let $\left(  A_{1},A_{2}\right)  $ be a TPP of $A.$ Then there
are irreducible component sets $\left\{  M_{i}\right\}  $ and $\left\{
N_{j}\right\}  $ for $A_{1}$ and $A_{2}$ respectively. By \textit{Corollary
3.1 }to \textit{Lemma 3.3} we can choose a synchronic basis$\left\{
x_{ji}\right\}  $associated with $\left\{  M_{i}\right\}  $ and $\left\{
N_{j}\right\}  .$ Now fix a pair of index $(i_{0},j_{0})$ and take
$W_{1}=M_{i_{0}},W_{2}=N_{j_{0}}.$ By definition $\left\{  x_{ji_{0}%
}|j=1,2,\cdots\right\}  $ and $\left\{  x_{j_{0}i}|i=1,2,\cdots\right\}  $ are
bases of $W_{1}$ and $W_{2}$ respectively. Thus we can define a bilinear map
$\otimes$ from $W_{1}\times W_{2}$ to $W$ such that $x_{ji_{0}}\otimes x_{j_{0}%
i}=x_{ji}.$ We claim that $(W_{1},W_{2},\otimes)$ is a TPS of $W$ associated
with $\left(  A_{1},A_{2}\right)  .$ As $\left\{  x_{ji_{0}}\otimes x_{j_{0}%
i}\right\}  =\left\{  x_{ji}\right\}  $ is a basis of $W,$ $(W_{1}%
,W_{2},\otimes)$ is obviously a TPS of $W.$ According to the definition, to
prove that it is a TPS associated with $\left(  A_{1},A_{2}\right)  $ we need
to show that
\begin{align*}
a\left(  x_{ji_{0}}\otimes x_{j_{0}i}\right)   &  =\left(  ax_{ji_{0}}\right)
\otimes x_{j_{0}i},\text{ }\forall a\in A_{1},\\
b\left(  x_{ji_{0}}\otimes x_{j_{0}i}\right)   &  =x_{ji_{0}}\otimes\left(
bx_{j_{0}i}\right)  ,\text{ }\forall b\in A_{2}.
\end{align*}
In fact, if%
\[
ax_{ji_{0}}=%
{\textstyle\sum_{k}}
x_{ki_{0}}a_{kj},\text{ }a_{kj}\in%
\Bbb{C} ,
\]
then%
\[
ax_{ji}=%
{\textstyle\sum_{k}}
x_{ki}a_{kj},\text{ }\forall i
\]
since $\left\{  x_{ji}\right\}  $ is a standard basis. It then follows that%
\[
a\left(  x_{ji_{0}}\otimes x_{j_{0}i}\right)  =ax_{ji}=%
{\textstyle\sum_{k}}
x_{ki}a_{kj}%
\]
and%
\[
\left(  ax_{ji_{0}}\right)  \otimes x_{j_{0}i}=\left(
{\textstyle\sum_{k}}
x_{ki_{0}}a_{kj}\right)  \otimes x_{j_{0}i}=%
{\textstyle\sum_{k}}
x_{ki}a_{kj}.
\]
This proves that $a\left(  x_{ji_{0}}\otimes x_{j_{0}i}\right)  =\left(
ax_{ji_{0}}\right)  \otimes x_{j_{0}i}.$ The other equation can be proved in
the same way.

We observe that in the finite dimensional case, if $\left(  A_{1}%
,A_{2}\right)  $ is a TPP of $A,$ then we have $A=$ $A_{1}\otimes A_{2}$ as a
direct consequence of the above theorem. This justifies calling $\left(
A_{1},A_{2}\right)  $ a TPP of $A.$

\ \ \ \ \ \ \ \ \ \ \ \ \ \ \ \ \ \ \ \ \ \ \ \ \ \ \ \ \ \ \ \ \ \ \ \ \ \ \ \ \ \ \ $\square
$

\textit{Remark} Theorem 4.1, together with the second half of Proposition 4.1,
provides a simple proof for Proposition 3.2.

Now we consider to what extent a given TPP determines the TPS associated with it.

\textbf{Definition 5.2.} \textit{Two tensor product structures }$(U_{1}%
,U_{2},\otimes_{1})$\textit{\ and }$(W_{1},W_{2},\otimes_{2})$\textit{\ of
}$W$\textit{\ are called equivalent if at least one of the following two
conditions is satisfied: (1) There are vector space isomorphisms }$\varphi
_{1}$\textit{: }$U_{1}\longrightarrow W_{1},$\textit{\ }$\varphi_{2}$\textit{:
}$U_{2}\longrightarrow W_{2},$\textit{\ and a complex number }$\alpha
$\textit{\ such that}%
\[
u_{1}\otimes_{1}u_{2}=\alpha\left(  \varphi_{1}u_{1}\otimes_{2}\varphi
_{2}u_{2}\right)  ,\text{ }\forall u_{1}\in U_{1},\text{ }u_{2}\in U_{2};
\]
\textit{(2) There are vector space isomorphisms }$\varphi_{1}$\textit{:
}$U_{1}\longrightarrow W_{2},$\textit{\ }$\varphi_{2}$\textit{: }%
$U_{2}\longrightarrow W_{1},$\textit{\ and a complex number }$\alpha
$\textit{\ such that}%
\[
u_{1}\otimes_{1}u_{2}=\alpha\left(  \varphi_{2}u_{2}\otimes_{2}\varphi
_{1}u_{1}\right)  ,\forall u_{1}\in U_{1},u_{2}\in U_{2}.
\]
\textit{The equivalent class of the TPS }$(W_{1},W_{2},\otimes)$\textit{\ is
denoted by }$\overline{(W_{1},W_{2},\otimes)}$\textit{, and the set of
equivalent classes of tensor product structures of }$W$\textit{\ is denoted by
}$T\left(  W\right)  .$

\textbf{Proposition 5.2} \textit{If two tensor product structures are
associated with the same TPP, then they are equivalent.}

\textit{Proof. }Let $(W_{1},W_{2},\otimes)$ be the TPS defined in the proof of
Theorem 3.1. It is then sufficient to show that an arbitrary TPS $(U_{1}%
,U_{2},\otimes_{1})$ associated with the TPP $\left(  A_{1},A_{2}\right)  $ is
equivalent to $(W_{1},W_{2},\otimes).$

By Lemma 5.1, $U_{1},U_{2}$ are irreducible modules. So $U_{1},U_{2}$ are
isomorphic to $W_{1},W_{2}$ as $A_{1}$ and $A_{2}$ modules respectively. It
then follows that there exist isomorphisms $\varphi_{1}$: $U_{1}%
\longrightarrow W_{1},$ $\varphi_{2}$: $U_{2}\longrightarrow W_{2},$ such that%
\[
a\cdot\varphi_{1}=\varphi_{1}\cdot a,\text{ }b\cdot\varphi_{2}=\varphi
_{2}\cdot b,\text{ }\forall a\in A_{1},\text{ }b\in A_{2}.
\]
Now fix a standard complete set of operators $\left\{  r,s\right\}  $ such
that $rx_{ji}=\lambda_{j}x_{ji},$ $sx_{ji}=\mu_{i}x_{ji}.$ By definition,
$W_{1}=M_{i_{0}},W_{2}=N_{j_{0}}$ and $x_{ji_{0}}\otimes x_{j_{0}i}=x_{ji}.$
As $r\in A_{1},$ $s\in A_{2}$, we then have%
\begin{align*}
r\left(  \varphi_{1}x_{ji_{0}}\otimes_{1}\varphi_{2}x_{j_{0}i}\right)   &
=\lambda_{j}\left(  \varphi_{1}x_{ji_{0}}\otimes_{1}\varphi_{2}x_{j_{0}%
i}\right)  ,\\
s\left(  \varphi_{1}x_{ji_{0}}\otimes_{1}\varphi_{2}x_{j_{0}i}\right)   &
=\mu_{i}\left(  \varphi_{1}x_{ji_{0}}\otimes_{1}\varphi_{2}x_{j_{0}i}\right)
,
\end{align*}
namely, $\left(
\varphi_{1}x_{ji_{0}}\otimes\varphi_{2}x_{j_{0}i}\right)  $
belongs to the same joint eigenspace of $\left\{  r,s\right\}  $
as $x_{ji}.$ But the joint eigenspaces of $\left\{  r,s\right\}  $
are all one dimensional, so we conclude that for each pair of
index $\left(  j,i\right)  $ there exists a complex number
$\alpha_{ji}$ such that
\[
x_{ji_{0}}\otimes x_{j_{0}i}=\alpha_{ji}\left(  \varphi_{1}x_{ji_{0}}%
\otimes_{1}\varphi_{2}x_{j_{0}i}\right)  .
\]
To prove the proposition we have to show that $\alpha_{ji}$ is independent of
$\left(  j,i\right)  .$

For different indice $j_{1},j_{2},$ take $a\in A_{1}$ such that $ax_{j_{1}%
i_{0}}=x_{j_{1}i_{0}}+x_{j_{2}i_{0}}.$ Note that the existence of such $a$ is
guaranteed by the fact that $A_{1}|_{M_{i_{0}}}=End(M_{i_{0}}).$ We then have%
\begin{align*}
a\left(  x_{j_{1}i_{0}}\otimes x_{j_{0}i}\right)   &  =\alpha_{j_{1}i}a\left(
\varphi_{1}x_{j_{1}i_{0}}\otimes_{1}\varphi_{2}x_{j_{0}i}\right)  ,\\
\left(  ax_{j_{1}i_{0}}\right)  \otimes x_{j_{0}i}  &  =\alpha_{j_{1}i}\left(
a\varphi_{1}x_{j_{1}i_{0}}\right)  \otimes_{1}\left(  \varphi_{2}x_{j_{0}%
i}\right)  =\alpha_{j_{1}i}\left(  \varphi_{1}ax_{j_{1}i_{0}}\right)
\otimes_{1}\left(  \varphi_{2}x_{j_{0}i}\right)  ,\\
\left(  x_{j_{1}i_{0}}+x_{j_{2}i_{0}}\right)  \otimes x_{j_{0}i}  &
=\alpha_{j_{1}i}\left(  \varphi_{1}\left(  x_{j_{1}i_{0}}+x_{j_{2}i_{0}%
}\right)  \right)  \otimes_{1}\left(  \varphi_{2}x_{j_{0}i}\right)  .
\end{align*}
Therefore,%
\begin{align*}
\alpha_{j_{1}i}\left(  \varphi_{1}x_{j_{1}i_{0}}\otimes_{1}\varphi_{2}%
x_{j_{0}i}\right)  +\alpha_{j_{2}i}\left(  \varphi_{1}x_{j_{2}i_{0}}%
\otimes_{1}\varphi_{2}x_{j_{0}i}\right)   &  =\alpha_{j_{1}i}\left(
\varphi_{1}x_{ji_{0}}\otimes_{1}\varphi_{2}x_{j_{0}i}\right)  +\alpha_{j_{1}%
i}\left(  \varphi_{1}x_{j_{2}i_{0}}\otimes_{1}\varphi_{2}x_{j_{0}i}\right)
,\\
\alpha_{j_{2}i}\left(  \varphi_{1}x_{j_{2}i_{0}}\otimes_{1}\varphi_{2}%
x_{j_{0}i}\right)   &  =\alpha_{j_{1}i}\left(  \varphi_{1}x_{j_{2}i_{0}%
}\otimes_{1}\varphi_{2}x_{j_{0}i}\right)  .
\end{align*}
It follows directly that $\alpha_{j_{1}i}=\alpha_{j_{2}i}.$ In the
same way, we can prove that $\alpha_{ji_{1}}=\alpha_{ji_{2}}$ for
different indices $i_{1},i_{2}.$ Consequently, all $\alpha_{ji}$'s
are equal. The proposition is thus proved.

\ \ \ \ \ \ \ \ \ \ \ \ \ \ \ \ \ \ \ \ \ \ \ \ \ \ \ \ \ \ \ \ \ \ \ \ \ \ \ \ \ \ \ \ \ \ \ \ \ \ \ \ \ \ \ \ \ $\square
$

\textbf{Definition 5.3} \textit{Two tensor product partitions
}$\left( A_{1},A_{2}\right)  $\textit{\ and }$\left(
B_{1},B_{2}\right) $\textit{\ are called equivalent if }$\left(
A_{1},A_{2}\right)  =\left( B_{1},B_{2}\right)  $\textit{\ or
}$\left(  A_{1},A_{2}\right)  =\left(
B_{2},B_{1}\right)  .$\textit{\ The equivalent class of }$\left(  A_{1}%
,A_{2}\right)  $\textit{\ is denoted by }$\overline{\left(  A_{1}%
,A_{2}\right)  },$\textit{\ and the set of equivalent classes of tensor
product partitions of }$End(W)$\textit{\ is denoted by }$P(W).$

\textbf{Lemma 5.2} \textit{TPS's associated with equivalent TPP's
are equivalent.}

\textit{Proof.} Let $\left(  A_{1},A_{2}\right)  $ and $\left(  B_{1}%
,B_{2}\right)  $ be equivalent TPP's. If $\left(
A_{1},A_{2}\right) =.\left(  B_{1},B_{2}\right)  ,$ then the
assertion is just what Proposition 4.2 says. Now suppose that
$\left(  A_{1},A_{2}\right) =\left(  B_{2},B_{1}\right)  .$ Let
$(U_{1},U_{2},\otimes_{1})$ and $(W_{1},W_{2},\otimes_{2})$ be
TPS's associated with $\left( A_{1},A_{2}\right)  $ and $\left(
B_{1},B_{2}\right)  $
respectively. We define a bilinear map $\otimes$: $U_{2}\times U_{1}%
\longrightarrow W$ \ such that $u_{2}\otimes u_{1}=u_{1}\otimes_{1}u_{2},$
$\forall u_{1}\in U_{1},u_{2}\in U_{2}.$ It is readily verified that
$(U_{2},U_{1},\otimes)$ is a TPS associated with $\left(  B_{1},B_{2}\right)
.$ It then follows from\textit{\ }Proposition 4.2 that there are vector space
isomorphisms $\varphi_{1}$: $W_{1}\longrightarrow U_{2},$ $\varphi_{2}$:
$W_{2}\longrightarrow U_{1},$ and a complex number $\alpha$ such that%
\[
W_{1}\otimes_{2}W_{2}=\alpha\left(  \varphi_{1}w_{1}\otimes\varphi_{2}%
w_{2}\right)  =\alpha\left(  \varphi_{2}w_{2}\otimes_{1}\varphi_{1}%
w_{1}\right)  ,\text{ }\forall w_{1}\in W_{1},\text{ }w_{2}\in W_{2}.
\]
This means, according to Definition 4.2, that $(U_{1},U_{2},\otimes_{1})$ and
$(W_{1},W_{2},\otimes_{2})$ are equivalent.

\ \ \ \ \ \ \ \ \ \ \ \ \ \ \ \ \ \ \ \ \ \ \ \ \ \ \ \ \ \ \ \ \ \ \ \ \ \ \ \ \ \ $\square
$

\textbf{Lemma 5.3} \ \textit{An arbitrary TPS of }$W$\textit{\ is equivalent
to a TPS of the form }$(W_{1},W_{2},\otimes)$\textit{\ with }$W_{1}%
,W_{2}\subseteq W.$

\textit{Proof.} Let $(V_{1},V_{2},\otimes_{1})$ be an arbitrary TPS of $W,$
and $\left\{  x_{i}\right\}  ,$ $\left\{  y_{j}\right\}  $ be respective bases
of $V_{1},V_{2}.$ Define two subspaces $W_{1},W_{2}$ of $W$ as%
\begin{align*}
W_{1} &  =V_{1}\otimes_{1}y_{1}\triangleq\left\{  u\otimes_{1}y_{1}|u\in
V_{1}\right\}  ,\\
W_{2} &  =x_{1}\otimes_{1}V_{2}\triangleq\left\{  x_{1}\otimes_{1}v|v\in
V_{2}\right\}  ,
\end{align*}
and a bilinear map $\otimes$: $W_{1}\times W_{2}\longrightarrow W$ such that
\[
\left(  x_{i}\otimes_{1}y_{1}\right)  \otimes\left(  x_{1}\otimes_{1}%
y_{j}\right)  =x_{i}\otimes_{1}y_{j}.
\]
It is then straightforward to check that $(W_{1},W_{2},\otimes)$ meets the requirement.

Let $\tau$ denote the map from the set of TPP's of $End(W)$ to the
set of TPS"s of $W$ under which a TPP is sent to a TPS associated
it. By Lemma 5.2\textit{\ }$\tau$ induces a map
$\overline{\tau}$ from $\mathcal{P}(W)$ to $\mathcal{T}\left(  W\right)  :$%
\[
\overline{\tau}\overline{\left(  A_{1},A_{2}\right)  }=\overline{\tau\left(
A_{1},A_{2}\right)  }.
\]
It follows from Proposition 5.1 and\textit{\ }Lemma 5.3 that $\overline{\tau}$
is surjective. Later we will prove that it is also injective.

\ \ \ \ \ \ \ \ \ \ \ \ \ \ \ \ \ \ \ \ \ \ \ \ \ \ \ \ \ \ \ \ \ \ \ \ \ \ \ \ \ \ \ \ \ \ \ \ \ \ \ \ \ \ \ \ \ \ \ $\square
$

\section{Inner Product Compatible Tensor Product Structure}

Up to now we have not taken into accounts the inner product
structure of $W.$ But in quantum mechanics, a physical space of
quantum states should be endowed with a reasonable inner product
so that the probability explanation of wave function could make
sense. In this section we proceed along to study the TPS in
connection with the inner product structure. We will introduce a
natural compatibility condition between these two structures as
the starting point. Further study will still be developed in the
context of TPP. Through out this section $W$ stands for a vector
space with the inner product $<$ $,$ $>,$ and $W_{1},W_{2}$ stand
for subspaces of $W.$

\textbf{Definition 6.1.} \textit{A TPS }$\left(  W_{1},W_{2},\otimes\right)
$\textit{\ of }$W$\textit{\ is called compatible with the inner product }%
$<$\textit{\ }$,$\textit{\ }$>$\textit{\ if }$\left\langle u_{1}\otimes
u_{2},v_{1}\otimes v_{2}\right\rangle =\left\langle u_{1},v_{1}\right\rangle
\left\langle u_{2},v_{2}\right\rangle $\textit{\ for all }$u_{1},v_{1}\in
W_{1}$\textit{\ and }$u_{2},v_{2}\in W_{2}.$

\textbf{Definition 6.2.}\textit{\ Let }$V_{1},V_{2}$\textit{\ be two modules
with the inner products }$<$\textit{\ }$,$\textit{\ }$>_{1}$\textit{and }%
$<$\textit{\ }$,$\textit{\ }$>_{2}$\textit{respectively. }$V_{1},V_{2}%
$\textit{\ are called }$U-$\textit{homomorphic (isomorphic) if
there exists an inner product preserving module homomorphism
(isomorphism) }$f$\textit{\ from }$V_{1}$\textit{\ to
}$V_{2}:\left\langle fv_{1},fv_{2}\right\rangle _{2}=\left\langle
v_{1},v_{2}\right\rangle _{1},$\textit{\ for all }$v_{1}\in
$\textit{\ }$V_{1},v_{2}\in V_{2}.$\textit{Such a }$f$\textit{\
will be called a }$U-$\textit{homomorphism (isomorphism) of
module.}

\textbf{Definition 6.3} \textit{A TPP }$\left(  A_{1},A_{2}\right)
$\textit{\ of }$End(W)$\textit{\ is called compatible with the
inner product }$<$\textit{\ }$,$\textit{\ }$>$\textit{\ if there
exist irreducible component sets }$\left\{  M_{i}\right\}
,$\textit{\ }$\left\{  N_{j}\right\} $\textit{\ of }$\left(
A_{1},A_{2}\right)  $\textit{\ such that (1) different
}$M_{i}$\textit{'s (}$N_{j}$\textit{'s) are orthogonal to one
another with
respect to }$<$\textit{\ }$,$\textit{\ }$>;$\textit{\ (2) all }$M_{i}%
$\textit{'s (}$N_{j}$\textit{'s) are }$U-$\textit{isomorphic }$A_{1}%
$\textit{(}$A_{2}$\textit{) modules.}

\textbf{Lemma 6.1. }\textit{A TPP }$\left(  A_{1},A_{2}\right)  $\textit{\ of
}$End(W)$\textit{\ is compatible with the inner product }$<$\textit{\ }%
$,$\textit{\ }$>$\textit{\ if and only if there exist irreducible component
sets of }$\left(  A_{1},A_{2}\right)  $\textit{\ which have an orthonormal
standard basis with respect to }$<$\textit{\ }$,$\textit{\ }$>.$

\textit{Proof}. If $\left(  A_{1},A_{2}\right)  $ is compatible
with the inner product, then there exist irreducible component
sets $\left\{ M_{i}\right\}  $ for $A_{1}$, all $M_{i}$'s being
$U-$isomorphic. Let $f_{i}$ be the inner product preserving
isomorphism from $M_{1}$ to $M_{i}$. Take an orthonormal basis
$\left\{  x_{j1}\right\}  $ of $M_{1}$, and define $x_{ji}=$
$f_{i}x_{j1}$. It is evident that $%
{\textstyle\bigcup_{i}}
\left\{  x_{ji}|j=1,2,\cdots\right\}  $ is an orthonormal
synchronic basis with respect to $\left\{  M_{i}\right\} $. Now
define $N_{j}$ to be the vector space spanned by $\left\{
x_{ji}|i=1,2,\cdots\right\} $, it then follows from Lemma 2.3 that
$\left\{  x_{ji}\right\}  $ is an orthonormal standard basis
associated with $\left\{  M_{i}\right\}  ,\left\{ N_{j}\right\} .$
This proves the necessity. For the sufficiency, just observe that
if $\left\{  x_{ji}\right\}  $ is an orthonormal standard basis
associated with $\left\{  M_{i}\right\}  ,\left\{ N_{j}\right\} ,$
then the linear maps $f_{i}:M_{1}\longrightarrow M_{i}$ which
sends $x_{j1}$ to $x_{ji}$ and $g_{j}:N_{1}\longrightarrow N_{j}$
which sends $x_{1i}$ to $x_{ji}$ are $U-$isomorphisms of module.

\ \ \ \ \ \ \ \ \ \ \ \ \ \ \ \ \ \ \ \ \ \ \ \ \ \ \ \ \ \ \ \ \ \ \ \ \ $\square
$

\textit{Remark} One easily sees from the proof of Lemma 5.1 that in Definition
5.3 the two conditions for $\left\{  M_{i}\right\}  $\ and the two conditions
for $\left\{  N_{j}\right\}  $\ are not independent. They actually imply each other.

We now probe the relation between the inner product compatible TPS
of $W$ and the inner product compatible TPP of $End(W).$

\textbf{Proposition 6.1} \textit{Let }$\left(  A_{1},A_{2}\right)
$\textit{\ be a TPP of }$End(W),$\textit{\ }$\left(  W_{1},W_{2}%
,\otimes\right)  $\textit{\ a TPS of }$W$\textit{\ associated with
}$\left( A_{1},A_{2}\right)  .$\textit{\ If }$\left(
W_{1},W_{2},\otimes\right) $\textit{\ is compatible with the inner
product, then so is }$\left( A_{1},A_{2}\right)  .$

\textit{Proof.} Let $\left\{  x_{j}\right\}  ,$ $\left\{  y_{i}\right\}  $ be
orthonormal bases of $W_{1}$ and $W_{2}$ respectively and define $M_{i}%
=W_{1}\otimes y_{i},$ $N_{j}=x_{j}\otimes W_{2}$. It follows that
$\left\{ M_{i}\right\}  ,\left\{  N_{j}\right\}  $ are irreducible
component sets of $\left(  A_{1},A_{2}\right)  $ and $\left\{
x_{j}\otimes y_{i}\right\}  $ is a standard basis associated with
them. If $\left(  W_{1},W_{2},\otimes\right) $ is compatible with
the inner product, then $\left\{  x_{j}\otimes y_{i}\right\}  $ is
an orthonormal basis of $W$ and hence an orthonormal standard
basis associated with $\left\{  M_{i}\right\}  ,\left\{
N_{j}\right\}  .$ The proposition thus follows from \textit{Lemma
6.1.}

Conversely, we have the following result.

\textbf{Proposition 6.2} If $\left(  A_{1},A_{2}\right)  $ is an
inner product compatible TPP of $End(W),$ then there exists an
inner product compatible TPS of $W$ associated with it.

\textit{Proof. }According to Lemma 5.1, we can choose irreducible
component sets $\left\{  M_{i}\right\}  ,\left\{  N_{j}\right\}  $
of $W$ and an orthonormal standard basis $\left\{  x_{ji}\right\}
$ associated with them. Fix a pair of index $(i_{0},j_{0})$, take
$W_{1}=M_{i_{0}},W_{2}=N_{j_{0}}$ and define a bilinear map
$\otimes$ from $W_{1}\times W_{2}$ to $W$ such that
$x_{ji_{0}}\otimes x_{j_{0}i}=x_{ji},$ as in the proof of Theorem
3.1. We can then check that $\left(  W_{1},W_{2},\otimes\right)  $
is a desired TPS of $W. $ That $\left(  W_{1},W_{2},\otimes\right)
$ is a TPS associated with $\left(  A_{1},A_{2}\right)  $ has been
proved in Theorem 3.1. It remains to show that for all
$u_{1},v_{1}\in W_{1}$ and $u_{2},v_{2}\in W_{2},$
\[
\left\langle u_{1}\otimes u_{2},v_{1}\otimes v_{2}\right\rangle =\left\langle
u_{1},v_{1}\right\rangle \left\langle u_{2},v_{2}\right\rangle .
\]
But this is an immediate consequence of the sequi-linearity of the inner
product $<$ $,$ $>$ and the relation:%
\[
\left\langle x_{j_{1}i_{0}}\otimes x_{j_{0}i_{1}},x_{j_{2}i_{0}}\otimes
x_{j_{0}i_{2}}\right\rangle =\left\langle x_{j_{1}i_{0}},x_{j_{2}i_{0}%
}\right\rangle \left\langle x_{j_{0}i_{1}},x_{j_{0}i_{2}}\right\rangle ,
\]
which follows from the assumption that $\left\{  x_{ji}\right\}  $ is an
orthonormal basis of $W.$

Next we turn to the problem of characterizing inner product
compatible TPP.

\textbf{Definition 6.4} \textit{Let }$B$\textit{\ be an extended subalgebra of
}$End(W).$\textit{\ If for every element }$b$\textit{\ }$\in$\textit{\ }%
$B$\textit{\ whose adjoint operator exists we have }$b^{\ast}\in
B,$\textit{\ where as usual }$b^{\ast}$\textit{\ stands for the adjoint
operator of }$b,$\textit{\ then }$B$\textit{\ is called a quasi-star extended
subalgebra.}

\textbf{Proposition 6.3} \textit{Let }$\left(  A_{1},A_{2}\right)
$\textit{\ be a TPP of }$End(W).$\textit{\ If it is compatible
with the inner product, then both }$A_{1}$\textit{\ and
}$A_{2}$\textit{\ are quasi-star extended subalgebras of
}$End(W).$\textit{\ Conversely, if }$A_{1}$\textit{(or }$A_{2}
$\textit{) is a quasi-star extended subalgebra and there exists an
irreducible component set }$\left\{  M_{i}\right\}  $\textit{(or }$N_{j}%
$\textit{) for }$A_{1}$\textit{(or }$A_{2}$\textit{) such that
different }$M_{i}$\textit{'s (or }$N_{j}$\textit{'s) are
orthogonal to one another, then }$\ \left(  A_{1},A_{2}\right)
$\textit{\ is compatible with the inner product.}

\textit{Proof.} If $\left(  A_{1},A_{2}\right)  $ is compatible
with the inner product, according to Lemma 5.1, we can choose an
orthonormal synchronic basis $\left\{  x_{ji}\right\}  $ with
respect to some irreducible component
set $\left\{  M_{i}\right\}  $ for $A_{1}.$Let $a\in$ $A_{1}.$ We can write%
\[
ax_{ji}=\sum_{k}x_{ki}a_{kj},
\]
where $a_{kj}$ is a complex number independent of the index $i.$We observe
that the adjoint operator of $a$ exists if and only if $\left\{  j|a_{kj}%
\neq0\right\}  $ is a finite set for each $k,$ and in that case we have%
\[
a^{\ast}x_{ji}=\sum_{k}x_{ki}\overline{a_{jk}},
\]
where $\overline{a_{jk}}$ stands for the complex-conjugate number of $a_{jk}.
$The fact that $\left\{  x_{ji}\right\}  $ is a synchronic basis with respect
to $\left\{  M_{i}\right\}  $ and $A_{1}|_{M_{i}}=End(W)$ then implies that
$a^{\ast}$ is an element of $A_{1}.$Consequently, $A_{1}$ is a quasi-star
extended subalgebra. The same conclusion for $A_{2}$ can be proved similarly.

To prove the second half of the proposition, let $A_{1}$ be a quasi-star
extended subalgebra, $\left\{  M_{i}\right\}  $ an irreducible component set
for $A_{1}$ such that elements from different $M_{i}$'s are orthogonal to one
another. When $A_{2}$ is a quasi-star extended subalgebra, the argument is
similar. Now from the proof of Lemma 5.1, we notice that for our purpose it
suffices to show that there exists an orthonormal synchronic basis with
respect to $\left\{  M_{i}\right\}  .$ The existence of such a basis is proved
as follows.

Let $\left\{  x_{j1}\right\}  $ be an orthonormal basis of $M_{1}$ and $%
{\textstyle\bigcup_{i}}
\left\{  x_{ji}|j=1,2,\cdots\right\}  $ a synchronic basis with respect to
$\left\{  M_{i}\right\}  $. If $a\in$ $A_{1}$ and $a^{\ast}$ exists, then we
have $a^{\ast}$ $\in$ $A_{1}.$ As a result, we can write%
\[
ax_{ji}=\sum_{k}x_{ki}a_{kj},a^{\ast}x_{ji}=\sum_{k}x_{ki}b_{kj},
\]
where $a_{kj},b_{kj}$ are complex numbers independent of the index $i.$ But
$\left\{  x_{j1}\right\}  $ is an orthonormal basis of $M_{1}$, so we have
$b_{kj}=\overline{a_{jk}}.$ Now by the definition of adjoint operator we have%
\[
\left\langle a^{\ast}x_{ji},x_{li}\right\rangle =\left\langle x_{ji}%
,ax_{li}\right\rangle .
\]
It then follows that%
\[
\left\langle \sum_{k}x_{ki}\overline{a_{jk}},x_{li}\right\rangle =\left\langle
x_{ji},\sum_{k}x_{ki}a_{kl}\right\rangle .
\]
Now if we choose $a\in$ $A_{1}$ such that $a_{mn}=\delta_{ml}\delta_{nl},$
from this equation we obtain
\[
\left\langle x_{ji},x_{li}\right\rangle =0,\text{ for }l\neq j;
\]
and if we choose $a\in$ $A_{1}$ such that $a_{mn}=\delta_{mj}\delta_{nl},$ we
obtain%
\[
\left\langle x_{ji},x_{ji}\right\rangle =\left\langle x_{li},x_{li}%
\right\rangle .
\]
Let $\left\langle x_{ji},x_{ji}\right\rangle =\alpha_{i}^{2}$
($\alpha_{1}=1$ by definition ),
$y_{ji}=\frac{x_{ji}}{\alpha_{i}},$ it is then check
that $%
{\textstyle\bigcup_{i}}
\left\{  y_{ji}|j=1,2,\cdots\right\}  $ is an orthonormal synchronic basis
with respect to $\left\{  M_{i}\right\}  .$

Before going on to another topic, let us pause to investigate the finite
dimensional case. In this case, we have the following much better result.

\textbf{Theorem 6.1} \textit{When }$W$\textit{\ is finite dimensional,
}$\left(  A_{1},A_{2}\right)  $\textit{\ is a TPP of }$End(W)$%
\textit{\ compatible with the inner product if and only if (1) }$A_{1},A_{2}%
$\textit{\ are star subalgebras of }$End(W);$\textit{\ (2) }$\left[
A_{1},A_{2}\right]  =0$\textit{\ and }$A=A_{1}\vee A_{2}.$

\textit{Proof.} The necessity follows from the first half of Proposition 6.3
directly. For the sufficiency, according to the second half of Proposition
6.3, we need only to show that as $A_{1}$($A_{2}$) module $W$ possesses a
decomposition into an orthogonal sum of irreducible $A_{1}$($A_{2}$)
submodules. But this is a direct consequence of the condition that
$A_{1},A_{2}$ are star subalgebras of $End(W).$ Indeed, if $V\subseteq W$ is
an $A_{1}$($A_{2}$) submodule, then the orthogonal complement $V^{\bot}$ of
$V$ is also an $A_{1}$($A_{2}$) submodule and we have the orthogonal
decomposition: $W=$ $V+V^{\bot}.$ Repeating this procedure, we will obtain the
desired decomposition after finite steps because $W$ is finite dimensional.

Finally, for the completeness, we now investigate the concept of standard
complete set of observables, the counterpart of standard complete set of operators.

\textbf{Definition 6.5} \textit{A standard complete set of operators }$\left(
r,t\right)  $\textit{\ of }$End(W)$\textit{\ is called a standard complete set
of observables if (1) both }$r$\textit{\ and }$t$\textit{\ are self-adjoint
operators; (2) there exists a characteristic set }$\left\{  M_{i}\right\}
$\textit{(or }$\left\{  N_{j}\right\}  $\textit{) of }$r$\textit{\ ( or }%
$t$\textit{\ ) consisting of subspaces orthogonal to one another.}

\textit{Remark} If $W$\ is finite dimensional, then the condition (1) implies
the condition (2).

The results listed below are about the relation between standard
complete set of observables and inner product compatible TPP.
Proposition 6.5 is just a special instance of Proposition 4.3, \
and the other propositions can be proved by similar argument as
presented in Section 2. Here we would rather omit the proof to
avoid redundancy.

\textbf{Proposition 6.4} \textit{If }$\left(  r,t\right)
$\textit{\ is a standard complete set of observables of
}$End(W),$\textit{\ then there exists an inner product compatible
TPP }$\left(  A_{1},A_{2}\right)  $\textit{\ of }$End(W)$\textit{\
containing }$\left(  r,t\right)  .$

\textbf{Proposition 6.5} \textit{Let }$\left(  r,t\right)
$\textit{\ be a standard complete set of observables of
}$End(W),$\textit{\ }$\left( A_{1},A_{2}\right)  $\textit{\ and
}$\left(  B_{1},B_{2}\right) $\textit{\ two inner product
compatible tensor product partitions containing
}$\left(  r,t\right)  .$\textit{\ Then there exists an isomorphism }%
$\varphi\in End(W)$\textit{, diagonal with respect to the basis consisting of
common eigenvectors of }$r$\textit{\ and }$t,$\textit{\ such that }%
$B_{1}=\varphi\cdot A_{1}\cdot\varphi^{-1}$\textit{\ and }$B_{2}=\varphi\cdot
A_{2}\cdot\varphi^{-1}.$

\textbf{Definition 6.6} \textit{The complete sets of observables }$\left(
r,t\right)  $\textit{\ and }$\left(  r^{\prime},t^{\prime}\right)
$\textit{\ are called complementary if (1) }$M_{i}=M_{i}^{\prime}%
$\textit{\ and all }$M_{i}$\textit{'s are }$U-$\textit{isomorphic normal
}$S_{r,r^{\prime}}$\textit{\ modules or (2) }$N_{j}=N_{j}^{\prime}%
$\textit{\ and all }$N_{j}$\textit{'s are }$U-$\textit{isomorphic normal
}$S_{t,t^{\prime}}$\textit{\ modules.}

\textbf{Proposition 6.6} \textit{An inner product compatible TPP
contains complementary standard complete sets of observables.}

\textbf{Proposition 6.7} \textit{If }$\left(  r,t\right)
$\textit{\ and }$\left(  \widetilde{r},\widetilde{t}\right)
$\textit{\ are complementary standard complete sets of
observables, then there exists a unique inner product compatible
TPP }$\left(  A_{1},A_{2}\right)  $\textit{\ such that
}$r,\widetilde{r}\in$\textit{\ }$A_{1}$\textit{\ and
}$t,\widetilde{t}\in A_{2}. $

\section{Product Vector Set and Relativity of Quantum Entanglement}

Now we turn to consider the set of decomposable vectors related to a TPS.
\ Decomposable and indecomposable vectors correspond respectively to product
and entangled states in physics. So from physical point of view, it is
meaningful to study this topic.

\textbf{Definition7.1} \ \textit{Let }$\left(  V_{1},V_{2},\otimes\right)
$\textit{\ be a TPS of }$W.$\textit{\ Then }$\left(  V_{1},V_{2}%
,\otimes\right)  $\textit{\ is called nontrivial if }$\dim V_{1},\dim V_{2}%
>1$\textit{\ and the subset }$\left\{  u\otimes v|u\in V_{1},v\in
V_{2}\right\}  $\textit{\ of }$W$\textit{\ is called the decomposable vector
set related to it.}

\textbf{Definition 7.2} \textit{A nonempty subset }$S$\textit{\ of }%
$W$\textit{\ is called a product vector set if it is a decomposable vector set
related to some TPS }$\left(  V_{1},V_{2},\otimes\right)  $\textit{\ of }%
$W$\textit{, and a nontrivial one if }$\left(  V_{1},V_{2},\otimes\right)
$\textit{\ is nontrivial.}

\textit{Remark} A product vector set of $W$\ is a nontrivial one if and only
if it is a proper subset of $W.$

\textbf{Definition 7.3} \textit{Let }$S$\textit{\ be a product vector set,
}$w$\textit{\ an element of }$W.$\textit{\ If }$w$\textit{\ belongs to }%
$S,$\textit{\ }$w$\textit{\ is called a product vector. Otherwise, it is
called an entangled vector.}

\textit{Remark} According to the definition, when we call an element of $W$\ a
product vector or an entangled one we should have in mind a TPS. Rigorously
speaking, in the above definition, $w$\ should be called a product vector or
an entangled one with respect to the TPS related to which $S$\ is a
decomposable vector set. Indeed, whether an element is a product vector or not
strongly depends on what TPS is considered. This point will become clear as we proceed.

Obviously, every product vector set of $W$ contains a basis of $W.$
Conversely, we have the following result.

\textbf{Proposition 7.1} \textit{Every basis of }$W$\textit{\ can be extended
to a nontrivial product vector set if }$\dim W$\textit{\ is not a prime
number.}

\textit{Proof}. Let $\left\{  w_{k}\right\}  $ be a basis of $W.$ $\dim W$
being not a prime number, we can choose subspaces $W_{1},W_{2}$ of $W$ with
$\dim W_{1},\dim W_{2}>1$ and the respective bases $\left\{  x_{j}\right\}
,\left\{  y_{i}\right\}  $ such that there exists a bijective map $f:\left\{
x_{j}\right\}  \times\left\{  y_{i}\right\}  \longrightarrow\left\{
w_{k}\right\}  .$ Extend $f$ bilinearly to a map $\otimes:W_{1}\times
W_{2}\longrightarrow W.$ It is then readily check that $\left(  W_{1}%
,W_{2},\otimes\right)  $ is a TPS of $W$ with respect to which $w_{k}$ is a
product vector for each $k.$ This proves the proposition.

\textbf{Corollary 7.1} \textit{An arbitrary element of }$W$\textit{\ is a
product vector with respect to some TPS, which can be chosen to be nontrivial
if }$\dim W$\textit{\ is not a prime number.}

\textbf{Corollary 7.2} \textit{An arbitrary element of }$W$\textit{\ is a
entangled vector with respect to some (nontrivial) TPS if }$\dim
W$\textit{\ is not a prime number.}

\textit{Proof.} Keep the same notation as in the proof of Proposition 3.3. Let
$w\in W.$ Choose a basis $\left\{  w_{k}\right\}  $of $W$ such that
$w=w_{1}+w_{2}$ and choose a map $f$ such that
\[
f\left(  x_{1},y_{2}\right)  =w_{1},\text{ }f\left(  x_{2},y_{1}\right)
=w_{2}.
\]
Then $w=x_{1}\otimes y_{2}+x_{2}\otimes y_{1}$ is an entangled vector with
respect to the TPS $\left(  W_{1},W_{2},\otimes\right)  .$

Similarly, we can prove the following results.

\textbf{Proposition 7.1}$^{\prime}$ \textit{Let }$W$\textit{\ a
space with inner product. Every orthonormal basis of }$W$\textit{\
can be extended to a nontrivial decomposable vector set related to
an inner product compatible TPS if }$\dim W$\textit{\ is not a
prime number.}

\textbf{Corollary 7.1}$^{\prime}$ \textit{An arbitrary element of }%
$W$\textit{\ is a decomposable vector with respect to some inner
product compatible TPS, which can be chosen to be nontrivial if
}$\dim W$\textit{\ is not a prime number.}

\textbf{Corollary 7.2}$^{\prime}$ \textit{An arbitrary element of }%
$W$\textit{\ is a entangled vector with respect to some
(nontrivial) inner product compatible TPS if }$\dim W$\textit{\ is
not a prime number.}

We denote by $\mathcal{D}(W)$ the set of product vector sets of
$W,$ and denote by $\sigma$ the map that sends each TPS of $W$ to
the decomposable vector set related to it. We observe that
equivalent TPP's have the same decomposable vector set. So
$\sigma$ naturally induces a map
$\overline{\sigma}$ from $\mathcal{T}\left(  W\right)  $ to $\mathcal{D}(W):$%
\[
\overline{\sigma}\overline{\left(  V_{1},V_{2},\otimes\right)  }=\sigma\left(
V_{1},V_{2},\otimes\right)  .
\]
By definition $\overline{\sigma}$ is a surjective map. Thus $\overline{\sigma
}\cdot\overline{\tau}$ is a surjective map from $\mathcal{P}(W)$ to
$\mathcal{D}(W).$ A proof of the bijectivity of this map is now in order.

\textit{Remark} Lemma 5.3 tells us that as far as product vector set is
concerned considering only the TPS of the form $(W_{1},W_{2},\otimes)$\ with
$W_{1},W_{2}\subseteq W$\ does not cause any loss of generality.

\textbf{Lemma 7.1} \textit{If }$\left(  A_{1},A_{2}\right)  $\textit{\ and
}$\left(  B_{1},B_{2}\right)  $\textit{\ are two tensor product partitions
such that }$\overline{\sigma}\overline{\tau}\overline{\left(  A_{1}%
,A_{2}\right)  }=\overline{\sigma}\overline{\tau}\overline{\left(  B_{1}%
,B_{2}\right)  },$\textit{\ then }$\overline{\left(  A_{1},A_{2}\right)
}=\overline{\left(  B_{1},B_{2}\right)  }.$

\textit{Proof.} Let $(W_{1},W_{2},\otimes),$ $(W_{1}^{\prime},W_{2}^{\prime
},\otimes^{\prime})$ be tensor product structures of $W$ associated with
$\left(  A_{1},A_{2}\right)  $ and $\left(  B_{1},B_{2}\right)  $
respectively. Denote by $D$ and $D^{\prime}$ the decomposable vector sets
related to $(W_{1},W_{2},\otimes)$ and $(W_{1}^{\prime},W_{2}^{\prime}%
,\otimes^{\prime})$ respectively. Suppose that $D=D^{\prime}.$ We then have to
show that either $A_{1}=B_{1}$ and $A_{2}=B_{2}$ or $A_{1}=B_{2}$ and
$A_{2}=B_{1}.$ Take a basis $\left\{  x_{j}\right\}  $ of $W_{1}$ and a basis
$\left\{  y_{i}\right\}  $ of $W_{2}.$ We observe that when $W_{1}$ or $W_{2}$
is of one dimension the proof is trivial. So we exclude this case in the
following argument. Now let us proceed in steps as follows.

(1) $W_{1}\otimes y_{1}\subseteq W_{1}^{\prime}\otimes^{\prime}y_{1}^{\prime}
$ for some $y_{1}^{\prime}\in W_{2}^{\prime}$ or $W_{1}\otimes y_{1}\subseteq
x_{1}^{\prime}\otimes^{\prime}W_{2}^{\prime}$ for some $x_{1}^{\prime}\in
W_{1}^{\prime}.$ Consider the elements $x_{1}\otimes y_{1}$ and $x_{2}\otimes
y_{1}$ of $D.$ Since $D=D^{\prime}$ there exist $x_{1}^{\prime},x_{2}^{\prime
}\in W_{1}^{\prime}$ and $y_{1}^{\prime},y_{1}^{\prime\prime}\in W_{2}%
^{\prime}$ such that%
\[
x_{1}\otimes y_{1}=x_{1}^{\prime}\otimes^{\prime}y_{1}^{\prime},\text{ }%
x_{2}\otimes y_{1}=x_{2}^{\prime}\otimes^{\prime}y_{1}^{\prime\prime}.
\]
On the other hand, we have $x_{1}\otimes y_{1}+x_{2}\otimes y_{1}=\left(
x_{1}+x_{2}\right)  \otimes y_{1}\in D.$ This implies that $x_{1}^{\prime
}\otimes^{\prime}y_{1}^{\prime}+x_{2}^{\prime}\otimes^{\prime}y_{1}%
^{\prime\prime}\in D^{\prime},$ that is, $x_{1}^{\prime}\otimes^{\prime}%
y_{1}^{\prime}+x_{2}^{\prime}\otimes^{\prime}y_{1}^{\prime\prime}$ is of rank
one with respect to $(W_{1}^{\prime},W_{2}^{\prime},\otimes^{\prime}).$ It
then follows from Proposition 1.1 that either $\left\{  x_{1}^{\prime}%
,x_{2}^{\prime}\right\}  $ or $\left\{  y_{1}^{\prime},y_{1}^{\prime\prime
}\right\}  $ is linearly dependent. Notice that both $\left\{  x_{1}^{\prime
},x_{2}^{\prime}\right\}  $ and $\left\{  y_{1}^{\prime},y_{1}^{\prime\prime
}\right\}  $ cannot be linearly dependent since $\left\{  x_{1}^{\prime
}\otimes^{\prime}y_{1}^{\prime},x_{2}^{\prime}\otimes^{\prime}y_{1}%
^{\prime\prime}\right\}  $ is linearly independent. We thus conclude that
either there exist a linearly independent set $\left\{  x_{1}^{\prime}%
,x_{2}^{\prime}\right\}  \subseteq W_{1}^{\prime}$ and a nonzero element
$y_{1}^{\prime\prime}\in W_{2}^{\prime}$ such that
\[
x_{1}\otimes y_{1}=x_{1}^{\prime}\otimes^{\prime}y_{1}^{\prime},\text{ }%
x_{2}\otimes y_{1}=x_{2}^{\prime}\otimes^{\prime}y_{1}^{\prime},
\]
or there exist a linearly independent set $\left\{  y_{1}^{\prime}%
,y_{1}^{\prime\prime}\right\}  \subseteq W_{2}^{\prime}$ and a nonzero element
$x_{1}^{\prime}\in W_{1}^{\prime}$ such that
\[
x_{1}\otimes y_{1}=x_{1}^{\prime}\otimes^{\prime}y_{1}^{\prime},\text{ }%
x_{2}\otimes y_{1}=x_{1}^{\prime}\otimes^{\prime}y_{1}^{\prime\prime}.
\]

If the first case happens, we assert, that $W_{1}\otimes
y_{1}\subseteq W_{1}^{\prime}\otimes^{\prime}y_{1}^{\prime}$ for
some $y_{1}^{\prime}\in W_{2}^{\prime},$ and if the second case
happens, $W_{1}\otimes y_{1}\subseteq
x_{1}^{\prime}\otimes^{\prime}W_{2}^{\prime}$ for some
$x_{1}^{\prime}\in W_{1}^{\prime}.$ In fact, if $\dim W_{1}=2$ the
assertion is evidently true. If $\dim W_{1}>2$ consider the
element $x_{j}\otimes y_{1},j\neq1,2.$ Let $x_{j}\otimes
y_{1}=x_{j}^{\prime}\otimes^{\prime}y_{1}^{\prime\prime\prime}.$
As argued above either $\left\{
x_{1}^{\prime},x_{i}^{\prime}\right\}  $ or $\left\{
y_{1}^{\prime},y_{1}^{\prime\prime\prime}\right\}  $ is linearly
dependent, but not both. Now the assertion clearly reduces to the
claim that $\left\{
y_{1}^{\prime},y_{1}^{\prime\prime\prime}\right\}  $ and $\left\{
x_{1}^{\prime},x_{j}^{\prime}\right\}  $ are respectively linearly
dependent in the first and the second cases. Suppose, on the
contrary, that $\left\{ x_{1}^{\prime},x_{j}^{\prime}\right\}  $
is linearly dependent in the first case. Then there exists a
complex number $\alpha$ such that $x_{j}\otimes y_{1}=\alpha
x_{1}^{\prime}\otimes^{\prime}y_{1}^{\prime\prime\prime}.$ It then
follows that
\[
\left(  x_{2}+x_{j}\right)  \otimes y_{1}=x_{2}^{\prime}\otimes^{\prime}%
y_{1}^{\prime}+\alpha x_{1}^{\prime}\otimes^{\prime}y_{1}^{\prime\prime\prime
}.
\]
Since $\left\{  x_{1}^{\prime},x_{2}^{\prime}\right\}  $ and $\left\{
y_{1}^{\prime},y_{1}^{\prime\prime\prime}\right\}  $ are both linearly
independent, the right hand side is an element of rank 2 with respect to
$(W_{1}^{\prime},W_{2}^{\prime},\otimes^{\prime})$ according to Proposition
1.1. But the left hand side is an element of rank 1 with respect to
$(W_{1},W_{2},\otimes).$ This contradicts the assumption that $D=D^{\prime}.$
The second case can be dealt with in the same way.

In the following discussion, we assume that $W_{1}\otimes y_{1}\subseteq
W_{1}^{\prime}\otimes^{\prime}y_{1}^{\prime}$ for some $y_{1}^{\prime}\in
W_{2}^{\prime}.$ The subsequent argument then proves that $A_{1}=B_{1}$ and
$A_{2}=B_{2}.$ If the other case happens, then a similar argument will prove
that $A_{1}=B_{2}$ and $A_{2}=B_{1}.$

(2) $W_{1}\otimes y_{1}=W_{1}^{\prime}\otimes^{\prime}y_{1}^{\prime}$ for some
$y_{1}^{\prime}\in W_{2}^{\prime}.$ Let $x^{\prime}$ be an arbitrary element
of $W_{1}^{\prime}.$ It suffices to show that $x^{\prime}$ $\otimes^{\prime
}y_{1}^{\prime}\in W_{1}\otimes y_{1}.$ Actually, we have
\[
x_{1}^{\prime}\otimes^{\prime}y_{1}^{\prime}=x_{1}\otimes y_{1},\text{ }%
x_{2}^{\prime}\otimes^{\prime}y_{1}^{\prime}=x_{2}\otimes y_{1}.
\]
Then by a similar argument as presented above we can prove that $x^{\prime}$
$\otimes^{\prime}y_{1}^{\prime}=x\otimes y_{1}$ for some $x\in W_{1},$ that is
$x^{\prime}$ $\otimes^{\prime}y_{1}^{\prime}\in W_{1}\otimes y_{1}.$

(3) There is a basis $\left\{  x_{j}^{\prime}\right\}  $ of $W_{1}$ and a
nonzero element $y_{1}^{\prime}$ of $W_{2}^{\prime}$ such that $x_{j}\otimes
y_{1}=x_{j}^{\prime}\otimes^{\prime}y_{1}^{\prime}.$ This is a direct
consequence of (1) and (2).

(4) There exists a linearly independent subset $\left\{  y_{i}^{\prime
}\right\}  $ of $W_{2}^{\prime}$ such that $x_{1}\otimes y_{i}=x_{1}^{\prime
}\otimes^{\prime}y_{i}^{\prime}$ for each $i.$ When $i=1$ the conclusion has
been proved. When $i\neq1$ consider the element $x_{1}\otimes y_{i}%
+x_{1}\otimes y_{1}$ of $D$. Let $x_{1}\otimes y_{i}=x_{1}^{\prime\prime
}\otimes^{\prime}y_{i}^{\prime}.$ Then $x_{1}\otimes y_{i}+x_{1}\otimes
y_{1}=x_{1}^{\prime\prime}\otimes^{\prime}y_{i}^{\prime}+x_{1}^{\prime}%
\otimes^{\prime}y_{1}^{\prime}$ is an element of $D^{\prime}.$ It follows that
exactly one of the two sets $\left\{  x_{1}^{\prime},x_{1}^{\prime\prime
}\right\}  $ and $\left\{  y_{1}^{\prime},y_{i}^{\prime}\right\}  $ is
linearly dependent. Clearly, what we need to show is that $\left\{
x_{1}^{\prime},x_{1}^{\prime\prime}\right\}  $ is linearly dependent. If, on
the contrary, $\left\{  y_{1}^{\prime},y_{i}^{\prime}\right\}  $ is linearly
dependent, then we have $y_{i}^{\prime}=\alpha_{i}y_{1}^{\prime}$ for some
$\alpha_{i}\in%
\Bbb{C} .$ As a result,
\[
x_{1}\otimes y_{i}+x_{2}\otimes y_{1}=\alpha_{i}x_{1}^{\prime\prime}%
\otimes^{\prime}y_{1}^{\prime}+x_{2}^{\prime}\otimes^{\prime}y_{1}^{\prime
}=\left(  \alpha_{i}x_{1}^{\prime\prime}+x_{2}^{\prime}\right)  \otimes
^{\prime}y_{1}^{\prime}.
\]
This is a contradiction since the right hand side is an element of $D^{\prime
}$ but the left hand side is not an element of $D.$ Finally, it is evident
that $\left\{  y_{i}^{\prime}\right\}  $ is linearly independent since
$\left\{  x_{1}\otimes y_{i}\right\}  $ is linearly independent.

(5) $x_{j}\otimes y_{i}=x_{j}^{\prime\prime}\otimes^{\prime}y_{i}^{\prime}$
for each $i.$ We need only to consider the case where $i,$ $j\neq1.$ Let
$x_{j}\otimes y_{i}=x_{j}^{\prime\prime}\otimes^{\prime}y_{i}^{\prime\prime}.$
We have to show that $\left\{  y_{i}^{\prime},y_{i}^{\prime\prime}\right\}  $
is linearly dependent. If this is not the case, by considering the rank of the
element $x_{1}\otimes y_{i}+x_{j}\otimes y_{i}$ we can prove that $\left\{
x_{1}^{\prime},x_{j}^{\prime\prime}\right\}  $ is linearly dependent. It then
follows that $x_{j}\otimes y_{i}+x_{1}\otimes y_{1}$ belongs to $D^{\prime}.$
But when $i,$ $j\neq1,$ the element $x_{j}\otimes y_{i}+x_{1}\otimes y_{1}$
does not lie in $D.$ This contradicts the assumption that $D=D^{\prime}.$

(6) $x_{j}\otimes y_{i}=x_{j}^{\prime}\otimes^{\prime}y_{i}^{\prime}$ for each
$i.$ By (5) we have $x_{j}\otimes y_{i}=x_{j}^{\prime\prime}\otimes^{\prime
}y_{i}^{\prime}.$ Then by a similar argument as presented in the proof of (4)
we can show that $\left\{  x_{j}^{\prime},x_{j}^{\prime\prime}\right\}  $ is
linearly dependent. So we can write $x_{j}\otimes y_{i}=\alpha x_{j}^{\prime
}\otimes^{\prime}y_{i}^{\prime},$ $\alpha\in%
\Bbb{C} .$ Let us prove that $\alpha=1.$ If $i=1$ or $j=1,$ there
is nothing to prove. When $i,j\neq1,$ consider the element $\left(
x_{1}+x_{j}\right) \otimes\left(  y_{1}+y_{i}\right)  $ of $D.$ We
have
\[
\left(  x_{1}+x_{j}\right)  \otimes\left(  y_{1}+y_{i}\right)  =\left(
x_{1}^{\prime}+\alpha x_{j}^{\prime}\right)  \otimes^{\prime}y_{i}^{\prime
}+\left(  x_{1}^{\prime}+x_{j}^{\prime}\right)  \otimes^{\prime}y_{1}^{\prime
}.
\]
As $\left\{  y_{1}^{\prime},y_{i}^{\prime}\right\}  $ is linearly independent,
it then follows that $\left\{  x_{1}^{\prime}+\alpha x_{j}^{\prime}%
,x_{1}^{\prime}+x_{j}^{\prime}\right\}  $ is linearly dependent. Consequently,
$\alpha=1.$ This proves the original assertion.

(7) $A_{1}=B_{1}$ and $A_{2}=B_{2}.$ Let $M_{i}=W_{1}\otimes y_{i},$
$M_{i}^{\prime}=W_{1}^{\prime}\otimes^{\prime}y_{i}^{\prime},$ $x_{ji}%
=x_{j}\otimes y_{i},x_{ji}^{\prime}=x_{j}^{\prime}\otimes^{\prime}%
y_{i}^{\prime}.$ Then $\left\{  M_{i}\right\}  $ is an irreducible component
set for $A_{1}$ and $%
{\textstyle\bigcup_{i}}
\left\{  x_{ji}|j=1,2,\cdots\right\}  $ is a synchronic basis with respect to
$\left\{  M_{i}\right\}  .$ Indeed, we have obviously the direct sum
decomposition $W=%
{\textstyle\sum_{i}}
\oplus M_{i}.$ Moreover, since $aM_{i}=\left(  aW_{1}\right)  \otimes y_{i},$
$\forall a\in A_{1},$ all $M_{i}$'s are irreducible $A_{1}$ modules isomorphic
to $W_{1}$ and $A_{1}$ has the same matrix representation with respect to the
bases $\left\{  x_{ji}|j=1,2,\cdots\right\}  ,i=1,2,\cdots.$ For the same
reason, $\left\{  M_{i}^{\prime}\right\}  $ is an irreducible component set
for $B_{1}$ and $%
{\textstyle\bigcup_{i}}
\left\{  x_{ji}^{\prime}|j=1,2,\cdots\right\}  $ is a synchronic basis with
respect to $\left\{  M_{i}^{\prime}\right\}  .$ But it follows from (3) that
$M_{i}=$ $M_{i}^{\prime}$ and by (6) we have $x_{ji}=x_{ji}^{\prime}.$
Considering $A_{1}|_{M_{i}}=B_{1}|_{M_{i}}=End(M_{i}),$ we then conclude that
$A_{1}=B_{1},$ and hence $A_{2}=B_{2}$ since $A_{2}=A_{1}^{\prime},$
$B_{2}=B_{1}^{\prime}.$

We are now in a position to present the following result on the
relation among the TPP, the TPS and the product vector set.

\textbf{Theorem 7.1} $\overline{\tau}$\textit{\ and }$\overline{\sigma}%
$\textit{\ are both bijective.}

Proof. By Lemma 7.1, $\overline{\sigma}\cdot\overline{\tau}$ is injective. It
follows that $\overline{\tau}$ is injective. But we have proved that
$\overline{\tau}$ is surjective, it is therefore bijective. $\overline{\sigma
}$ is surjective by definition. For the bijectivity, just notice that
$\overline{\sigma}=\left(  \overline{\sigma}\cdot\overline{\tau}\right)
\cdot\overline{\tau}^{-1}.$ The claim then follows directly.

\textit{Remark} This theorem shows that each product vector set is
characterized by an unordered pair $\left\{  A_{1},A_{2}\right\}  $\ where
$\left(  A_{1},A_{2}\right)  $\ is a TPP.

In the remaining part of this section, we keep on studying the property of
product vector set.

\textbf{Proposition 7.2} \textit{Let }$D$\textit{\ be a subset of }%
$W.$\textit{\ Then }$D$\textit{\ is a product vector set if and only if there
exist a TPP }$\left(  A_{1},A_{2}\right)  $\textit{\ of }$End(W)$\textit{\ and
an nonzero element }$w$\textit{\ of }$W$\textit{\ such that }$D=A_{1}A_{2}%
w$\textit{\ and }$\left(  A_{1}w,A_{2}w,\otimes_{D}\right)  $\textit{\ is a
TPS of }$W,$\textit{\ where }$\otimes_{D}$\textit{\ is defined as }%
$aw\otimes_{D}bw=abw,$\textit{\ }$\forall a\in A_{1}$\textit{, }$b\in A_{2}%
$\textit{.}

\textit{Proof.} If $D$ is a product vector set, then there exist a TPP
$\left(  A_{1},A_{2}\right)  $ and a TPS $(W_{1},W_{2},\otimes)$ associated
with it such that
\[
D=\left\{  u\otimes v|u\in W_{1},v\in W_{2}\right\}  .
\]
Take nonzero elements $u\in W_{1}$, $v\in W_{2}$ and let $w=$
$u\otimes v.$ By Lemma 3.1 $W_{1},W_{2}$ are respectively
irreducible $A_{1},A_{2}$ modules. So $A_{1}u=W_{1}$,
$A_{2}v=W_{2}$ and hence $D=A_{1}A_{2}w.$ On the other hand, we
have by definition $aw\otimes_{D}bw=au\otimes bv.$ It then follows
directly that $\left(  A_{1}w,A_{2}w,\otimes_{D}\right)  $ is a
TPS of $W.$ The necessity is thus proved. For the sufficiency,
just observe that $D$ is
exactly the decomposable vector set related to the TPS $\left(  A_{1}%
w,A_{2}w,\otimes_{D}\right)  .$

Let $\varphi$ be an automorphism of $W$ and $\left(  A_{1},A_{2}\right)  $ a
TPP. Then $\left(  \varphi\cdot A_{1}\cdot\varphi^{-1},\varphi\cdot A_{2}%
\cdot\varphi^{-1}\right)  $ is also a TPP. Moreover $\varphi$ induces a map
$\overline{\varphi}$ from $\mathcal{P}(W)$ to $\mathcal{P}(W):\overline
{\varphi}\overline{\left(  A_{1},A_{2}\right)  }=\overline{\left(
\varphi\cdot A_{1}\cdot\varphi^{-1},\varphi\cdot A_{2}\cdot\varphi
^{-1}\right)  }.$

\textbf{Lemma 7.2} \textit{If }$\varphi$\textit{\ is an automorphism of }$W,
$\textit{\ then }$\overline{\sigma}\cdot\overline{\tau}\cdot\overline{\varphi
}=\varphi\cdot\overline{\sigma}\cdot\overline{\tau}.$

\textit{Proof.} Let $\left(  A_{1},A_{2}\right)  $ be a TPP and $(W_{1}%
,W_{2},\otimes)$ a TPS associated with it. We assert that $(\varphi
W_{1},\varphi W_{2},\otimes_{\varphi})$ is a TPS associated with $\left(
\varphi\cdot A_{1}\cdot\varphi^{-1},\varphi\cdot A_{2}\cdot\varphi
^{-1}\right)  ,$ where $\otimes_{\varphi}=\varphi\cdot\otimes\cdot\left(
\varphi^{-1}\times\varphi^{-1}\right)  :$%
\[
\left(  \varphi w_{1}\right)  \otimes_{\varphi}\left(  \varphi w_{2}\right)
=\otimes_{\varphi}\left(  \varphi w_{1},\varphi w_{2}\right)  =\varphi\left(
w_{1}\otimes w_{2}\right)  ,\text{ }\forall w_{1}\in W_{1},\text{ }w_{2}\in
W_{2}.
\]
$(\varphi W_{1},\varphi W_{2},\otimes_{\varphi})$ is obviously a TPS. On the
other hand, for all $w_{1}\in W_{1},$ $w_{2}\in W_{2},$ we have%
\begin{align*}
\left(  \varphi a\varphi^{-1}\right)  \left(  \left(  \varphi w_{1}\right)
\otimes_{\varphi}\left(  \varphi w_{2}\right)  \right)   &  =\varphi a\left(
w_{1}\otimes w_{2}\right)  =\varphi\left(  aw_{1}\otimes w_{2}\right) \\
&  =\left(  \varphi aw_{1}\right)  \otimes_{\varphi}\left(  \varphi
w_{2}\right)  =\left(  \left(  \varphi a\varphi^{-1}\right)  \left(  \varphi
w_{1}\right)  \right)  \otimes_{\varphi}\left(  \varphi w_{2}\right)  ,\forall
a\in A_{1},
\end{align*}
and similarly%
\[
\left(  \varphi b\varphi^{-1}\right)  \left(  \left(  \varphi w_{1}\right)
\otimes_{\varphi}\left(  \varphi w_{2}\right)  \right)  =\left(  \varphi
w_{1}\right)  \otimes_{\varphi}\left(  \left(  \varphi b\varphi^{-1}\right)
\left(  \varphi w_{2}\right)  \right)  ,\text{ }\forall b\in A_{2}.
\]
The assertion then follows. Now it is clearly seen that $\overline{\sigma
}\overline{\tau}\overline{\varphi}\overline{\left(  A_{1},A_{2}\right)
}=\varphi\overline{\sigma}\overline{\tau}\overline{\left(  A_{1},A_{2}\right)
}$ is a direct consequence of the definition of $\otimes_{\varphi}.$ This
proves the lemma.

\textbf{Lemma 7.3} \textit{Let }$\left(  r,t\right)  $\textit{\ be a standard
complete set of operators of }$A$\textit{, }$\left\{  M_{i}\right\}
$\textit{, }$\left\{  N_{j}\right\}  $\textit{\ the characteristic sets of
}$r$\textit{\ and }$t$\textit{\ respectively. If }$\left(  A_{1},A_{2}\right)
$\textit{\ is a TPP containing }$\left(  r,t\right)  ,$\textit{\ then we have
}$M_{i},N_{j}\subseteq\overline{\sigma}\overline{\tau}\overline{\left(
A_{1},A_{2}\right)  }.$

\textit{Proof.} According to Lemma 4.2, $\left\{  M_{i}\right\}  $
and $\left\{  N_{j}\right\}  $ are irreducible component sets for
$A_{1}$ and $A_{2} $ respectively. Let $\left\{  x_{ji}\right\}  $
be a synchronic basis associated with $\left\{  M_{i}\right\}  $
and $\left\{  N_{j}\right\}
.$ Then it follows from Theorem 5.1 that there exists a TPS $(W_{1}%
,W_{2},\otimes)$ associated with $\left(  A_{1},A_{2}\right)  $ such that for
some $i_{0},j_{0}$
\[
M_{i}=W_{1}\otimes x_{j_{0}i},\text{ }N_{j}=x_{ji_{0}}\otimes W_{2}.
\]
This implies that $M_{i},N_{j}\subseteq\overline{\sigma}\overline{\tau
}\overline{\left(  A_{1},A_{2}\right)  }.$

Now we conclude this section by the following results about the construction
of product vector set.

\textbf{Proposition 7.3} \textit{Let }$\left(  r,t\right)  $\textit{\ be a
standard complete set of operators of }$A\left(  =End(W)\right)  $\textit{,
}$\left\{  M_{i}\right\}  $\textit{, }$\left\{  N_{j}\right\}  $\textit{\ the
characteristic sets of }$r$\textit{\ and }$t$\textit{\ respectively, then
there exists a product vector set }$D$\textit{\ containing }$M_{i}%
$\textit{\ and }$N_{j}$\textit{\ as subsets. Moreover, if we require, in
addition, that for }$u\otimes v\in D$\textit{\ }%
\begin{equation}
r\left(  u\otimes v\right)  =\left(  ru\right)  \otimes v,\text{ }t\left(
u\otimes v\right)  =u\otimes\left(  tv\right)  ,\tag{(*)}%
\end{equation}
\textit{then such product vector set is unique up to an automorphism of }%
$W$\textit{, which is diagonal with respect to the basis consisting of common
eigenvectors of }$r$\textit{\ and }$t.$

\textit{Proof.} Take a TPP $\left(  A_{1},A_{2}\right)  $ containing $\left(
r,t\right)  $ and define $D=\overline{\sigma}\overline{\tau}\overline{\left(
A_{1},A_{2}\right)  }.$ Then $D$ meets the requirement according to Lemma 7.3.
This proves the first half of the proposition.

To prove the second half of the proposition, let $\left(  A_{1},A_{2}\right)
$ be a TPP and $(W_{1},W_{2},\otimes)$ a TPS associated with it such that
$M_{i},N_{j}\subseteq\overline{\sigma}\overline{\tau}\overline{\left(
A_{1},A_{2}\right)  }=\overline{\sigma}\overline{(W_{1},W_{2},\otimes)}.$ We
need to show that $\overline{\left(  A_{1},A_{2}\right)  }$ is determined
uniquely up to an automorphism of $W$.

First we recall that by the definition of standard complete set of operators,
there are two sets of distinct complex numbers $\left\{  \lambda_{j}\right\}
$ and $\left\{  \mu_{i}\right\}  $ and two decompositions of $W$ into direct
sum of subspaces
\[
W=\sum_{i}\oplus M_{i}=\sum_{j}\oplus N_{j}%
\]
such that
\[
M_{i}=\sum_{j}\oplus%
\Bbb{C}
x_{ji},\text{ }N_{j}=\sum_{i}\oplus%
\Bbb{C}
x_{ji}%
\]
where $x_{ji}$ is the common eigenvector of $r$ and $t$:
\[
rx_{ji}=\lambda_{j}x_{ji},\text{ }tx_{ji}=\mu_{i}x_{ji}.
\]
Now, let us proceed in steps.

(1) There exist a basis $\left\{  u_{j}\right\}  $ of $W_{1}$ and a basis
$\left\{  v_{i}\right\}  $ of $W_{2}$ such that $M_{i}=W_{1}\otimes v_{i}$ and
$N_{j}=u_{j}\otimes W_{2}.$ Since $M_{i}\subseteq\sigma(W_{1},W_{2},\otimes),$
there exist $u_{j}\in W_{1}$ and $v_{i_{j}}\in W_{2}$ such that $x_{ji}%
=u_{j}\otimes v_{i_{j}}.$Then by the condition (*) we have%
\begin{align*}
\lambda_{j}u_{j}\otimes v_{i_{j}}  &  =rx_{ji}=\left(  ru_{j}\right)  \otimes
v_{i_{j}},\\
\mu_{i}u_{j}\otimes v_{i_{j}}  &  =tx_{ji}=u_{j_{1}}\otimes\left(  tv_{i_{j}%
}\right)  ,
\end{align*}
and hence%
\[
ru_{j}=\lambda_{j}u_{j},\text{ }tv_{i_{j}}=\mu_{i}v_{i_{j}}.
\]
Now using the fact that $M_{i}$ is a vector space one can easily prove that
$v_{i_{j_{1}}},v_{i_{j_{2}}}$ are linearly dependent for different
$j_{1},j_{2}.$ Let $W_{1}^{\prime}$ be the subspace of $W_{1}$ spanned by
$\left\{  u_{j}\right\}  ,$ it then follows that there exists $v_{i}\in W_{2}$
such that $tv_{i}=$ $\mu_{i}v_{i}$ and $M_{i}=W_{1}^{\prime}\otimes v_{i}.$
Notice that $W_{1}\otimes v_{i}$ is included in the eigenspace of $t$
corresponding to the eigenvalue $\mu_{i},$ which is just $M_{i}.$ Thus we
have
\[
M_{i}=W_{1}^{\prime}\otimes v_{i}\subseteq W_{1}\otimes v_{i}\subseteq M_{i},
\]
hence $M_{i}=W_{1}\otimes v_{i}.$ Similarly, there exists $u_{j}$ $\in W_{1}$
such that $N_{j}=u_{j}\otimes W_{2}.$ Finally, it is easy to check that
$\left\{  u_{j}\right\}  ,\left\{  v_{i}\right\}  $ are bases of $W_{1}$ and
$W_{2}$ respectively.

(2) $\left(  A_{1},A_{2}\right)  $ is a TPP containing $\left(  r,t\right)
.$According to (1), $\left\{  M_{i}\right\}  ,\left\{  N_{j}\right\}  $ are
irreducible component sets of $\left(  A_{1},A_{2}\right)  ,$ and $\left\{
u_{j}\otimes v_{i}\right\}  $ is a standard basis associated with them. The
assertion then follows.

Finally, by Proposition 4.3, we conclude from (2) that $\left(  A_{1}%
,A_{2}\right)  $ is determined uniquely up to an automorphism of $W$. The
conclusion then follows from Lemma 7.2.

\textbf{Proposition 7.4} \textit{Let }$\left(  r,t\right)
$\textit{\ be a standard complete set of observables of }$A\left(
=End(W)\right)  $\textit{, }$\left\{  M_{i}\right\}  $\textit{,
}$\left\{  N_{j}\right\}  $\textit{\ the characteristic sets of
}$r$\textit{\ and }$t$\textit{\ respectively, then there exists an
inner product compatible TPS whose decomposable vector set
}$D$\textit{\ contains }$M_{i}$\textit{\ and }$N_{j}$\textit{\ as
subsets.
Moreover, if we require, in addition, that for }$u\otimes v\in D$\textit{\ }%
\[
r\left(  u\otimes v\right)  =\left(  ru\right)  \otimes v,\text{ }t\left(
u\otimes v\right)  =u\otimes\left(  tv\right)  ,
\]
\textit{then such decomposable vector set is unique up to an automorphism of
}$W$\textit{, which is diagonal with respect to the basis consisting of common
eigenvectors of }$r$\textit{\ and }$t.$

The proof of this proposition is similar to that of Proposition 7.3. We would
rather omit it.

\section{Examples for Relativity of Quantum Entanglement}

In this section, we will analyze three examples as an illustration of the
theory developed above. The first example concerns the so called Bell states,
the second one deals with entanglement in Bargmann space, and the third one is
about entanglement with respect to the coordinate of mass of center mentioned
in the introduction. In the subsequent discussion, following the physical
convention, we will sometimes call a vector a state.

\subsection{Entanglement of Bell States}

Now let us study the first example that has been considered
extensively, but not rigorously from mathematical point of view,
by Zanardi et al\cite{z1}.

Consider the system $S_{AB}$ consisting of two spin $\frac{1}{2}$
particles labelled by $A$ and $B$ respectively. We will not be
interested in the dependence of the wave functions on the
coordinates. For a spin $\frac{1}{2}$
particle, the spin operator $\overrightarrow{S}$ takes the form ($\hbar=1$)%
\begin{equation}
\overrightarrow{S}=\frac{1}{2}\overrightarrow{\sigma}=\frac{1}{2}(\sigma
_{x},\sigma_{y},\sigma_{z}),
\end{equation}
where%
\begin{equation}
\sigma_{x}=\left(
\begin{array}
[c]{cc}%
0 & 1\\
1 & 0
\end{array}
\right)  ,\sigma_{y}=\left(
\begin{array}
[c]{cc}%
0 & -i\\
i & 0
\end{array}
\right)  ,\sigma_{z}=\left(
\begin{array}
[c]{cc}%
1 & 0\\
0 & -1
\end{array}
\right)
\end{equation}
are Pauli matrices. Conventionally, the two eigenvectors of $S_{z}$ are
denoted by $\left\vert \uparrow\right\rangle ,\left\vert \downarrow
\right\rangle ,$ which belong to the eigenvalues $\frac{1}{2}$ and $-\frac
{1}{2}$ respectively. To distinguish different particles, for operators we
introduce an upper script and for states we introduce a lower script. For
example, $\sigma_{z}^{A}$ denotes the spin operator for particle $A$ and
$\left\vert \uparrow\right\rangle _{B}$ denotes the eigenvector of $\sigma
_{z}^{B}.$

Let $V_{1}$ be the vector space spanned by $\left\{  \left\vert
\uparrow \right\rangle _{A},\left\vert \downarrow\right\rangle
_{A}\right\}  $ and $V_{2}$ the vector space spanned by $\left\{
\left\vert \uparrow\right\rangle _{B},\left\vert
\downarrow\right\rangle _{B}\right\}  .$ Then the space of states
of the system $S_{AB},$ which we denote by $W,$ has a God given
inner product compatible TPS $\left(
V_{1},V_{2},\otimes_{0}\right)  :W$ is taken or defined to be the
vector space spanned by the linearly independent set
\[
\left\{  \left\vert \uparrow\right\rangle _{A}\otimes_{0}\left\vert
\uparrow\right\rangle _{B},\left\vert \uparrow\right\rangle _{A}\otimes
_{0}\left\vert \downarrow\right\rangle _{B},\left\vert \downarrow\right\rangle
_{A}\otimes_{0}\left\vert \uparrow\right\rangle _{B},\left\vert \downarrow
\right\rangle _{A}\otimes_{0}\left\vert \downarrow\right\rangle _{B}\right\}
,
\]
which is usually written as
\[
\left\{  \left\vert \uparrow\right\rangle _{A}\left\vert \uparrow\right\rangle
_{B},\left\vert \uparrow\right\rangle _{A}\left\vert \downarrow\right\rangle
_{B},\left\vert \downarrow\right\rangle _{A}\left\vert \uparrow\right\rangle
_{B},\left\vert \downarrow\right\rangle _{A}\left\vert \downarrow\right\rangle
_{B}\right\}  .
\]
We call this TPS God given because we are not able to define the above four
vectors definitely as elements of $W.$ As a matter of fact, they are tacitly
understood as the common eigenstates of $\sigma_{z}^{A}$ and $\sigma_{z}^{B}.$
But the problem is still there: the phase is not and cannot be determined. We
have seen that from mathematical point of view, the bilinear map $\otimes_{0}%
$is not well defined. But as far as physics is concerned, we have no choice
but take it for granted and make it the starting point of our discussions in
this paper.

The so called Bell states are defined as follows:%
\begin{align}
\left\vert \psi^{\pm}\right\rangle _{AB} &  =\frac{1}{\sqrt{2}}\left(
\left\vert \uparrow\right\rangle _{A}\left\vert \downarrow\right\rangle
_{B}\pm\left\vert \downarrow\right\rangle _{A}\left\vert \uparrow\right\rangle
_{B}\right)  ,\\
\left\vert \phi^{\pm}\right\rangle _{AB} &  =\frac{1}{\sqrt{2}}\left(
\left\vert \uparrow\right\rangle _{A}\left\vert \uparrow\right\rangle _{B}%
\pm\left\vert \downarrow\right\rangle _{A}\left\vert \downarrow\right\rangle
_{B}\right)  .\nonumber
\end{align}
Obviously, they are maximally entangled states with respect to the
TPS $\left(  V_{1},V_{2},\otimes_{0}\right)  .$ The Bell states
form an orthonormal basis of $W,$ so according to Proposition
7.1$^{\prime}$, there exists an inner product compatible TPS with
respect to which they are product states. Let us explicitly
construct such tensor product structures.

Let $R_{x}(\pi)$ be the two bi-particle rotation operator through the angle
$\pi$ about the $x-$axis. We have%
\begin{equation}
R_{x}(\pi)=\left(
\begin{array}
[c]{cc}%
0 & i\\
i & 0
\end{array}
\right)  \otimes_{0}\left(
\begin{array}
[c]{cc}%
0 & i\\
i & 0
\end{array}
\right)  .
\end{equation}
Notice that $R_{x}(\pi)$ is unitary and self-adjoint as well. It is easy to
check that
\begin{equation}
R_{x}(\pi)\left\vert \psi^{\pm}\right\rangle _{AB}=\mp\left\vert \psi^{\pm
}\right\rangle _{AB},\text{ }R_{x}(\pi)\left\vert \phi^{\pm}\right\rangle
_{AB}=\mp\left\vert \phi^{\pm}\right\rangle _{AB}.
\end{equation}
Denote by $M_{1},M_{2}$ the subspaces spanned by $\left\{  \left\vert
\psi^{\pm}\right\rangle _{AB}\right\}  $ and $\left\{  \left\vert \phi^{\pm
}\right\rangle _{AB}\right\}  $ respectively, and by $N_{1},N_{2}$ the
subspaces spanned by $\left\{  \left\vert \psi^{+}\right\rangle _{AB}%
,\left\vert \phi^{+}\right\rangle _{AB}\right\}  $ and $\left\{  \left\vert
\psi^{-}\right\rangle _{AB},\left\vert \phi^{-}\right\rangle _{AB}\right\}  $
respectively. Then we have the orthogonal decomposition%
\[
W=M_{1}\oplus M_{2}=N_{1}\oplus N_{2}.
\]
Now let $S_{z}=S_{z}^{A}+S_{z}^{B}.$ Clearly we have%
\begin{equation}
S_{z}^{2}\left\vert \psi^{\pm}\right\rangle _{AB}=0,\text{ }S_{z}%
^{2}\left\vert \phi^{\pm}\right\rangle _{AB}=\left\vert \phi^{\pm
}\right\rangle _{AB}.
\end{equation}
It then follows that $\left(  R_{x}(\pi),S_{z}^{2}\right)  $ is a standard
complete set of observables, and $\left\{  M_{1},M_{2}\right\}  ,$ $\left\{
N_{1},N_{2}\right\}  $ are characteristic sets of $R_{x}(\pi)$ and $S_{z}^{2}
$ respectively. According to Proposition 6.4, there exists an inner product
compatible TPS whose decomposable vector set contains $M_{1},M_{2}%
,N_{1},N_{2}.$

Let us first construct an inner product compatible TPP contains
$\left( R_{x}(\pi),S_{z}^{2}\right)  .$ Using the previous
notation, let $x_{11}=\left\vert \psi^{+}\right\rangle _{AB},$
$x_{21}=\left\vert \psi ^{-}\right\rangle _{AB},$
$x_{12}=\left\vert \phi^{+}\right\rangle _{AB},$
$x_{22}=\left\vert \phi^{-}\right\rangle _{AB}.$ Define the
subalgebras $A_{1},A_{2}\subseteq End(W)$ as follows. An element
$a$ belongs to $A_{1}$ if and only if
\begin{align}
a\left(  x_{11},x_{21}\right)   &  =\left(  x_{11},x_{21}\right)  \left(
\begin{array}
[c]{cc}%
a_{11} & a_{12}\\
a_{21} & a_{22}%
\end{array}
\right)  ,\\
a\left(  x_{12},x_{22}\right)   &  =\left(  x_{12},x_{22}\right)  \left(
\begin{array}
[c]{cc}%
a_{11} & a_{12}\\
a_{21} & a_{22}%
\end{array}
\right)  ;
\end{align}
and an element $b$ belongs to $A_{2}$ if and only if%
\begin{align}
b\left(  x_{11},x_{12}\right)   &  =\left(  x_{11},x_{12}\right)  \left(
\begin{array}
[c]{cc}%
b_{11} & b_{12}\\
b_{21} & b_{22}%
\end{array}
\right)  ,\\
b\left(  x_{21},x_{22}\right)   &  =\left(  x_{21},x_{22}\right)  \left(
\begin{array}
[c]{cc}%
b_{11} & b_{12}\\
b_{21} & b_{22}%
\end{array}
\right)  ,
\end{align}
where $a_{ij},b_{ij}$ are arbitrary complex numbers. It is straightforward to
check that $\left(  A_{1},A_{2}\right)  $ is the desired TPP and $\left\{
M_{1},M_{2}\right\}  ,$ $\left\{  N_{1},N_{2}\right\}  $ are irreducible
component sets of $\left(  A_{1},A_{2}\right)  .$

Next we construct an inner product compatible TPS associated with
$\left( A_{1},A_{2}\right)  .$ Take $W_{1}=M_{1},W_{2}=N_{2}$ and
define a bilinear
map $\otimes$ from $W_{1}\times W_{2}$ to $W$ such that%
\begin{align}
x_{11}\otimes x_{21} &  =x_{11},\text{ }x_{11}\otimes x_{22}=x_{12},\\
x_{21}\otimes x_{21} &  =x_{21},\text{ }x_{21}\otimes x_{22}=x_{22}.
\end{align}
Then $\left(  W_{1},W_{2},\otimes\right)  $ is an inner product
compatible
TPS. Notice that%
\begin{align}
M_{1} &  =W_{1}\otimes x_{21},\text{ }M_{2}=W_{1}\otimes x_{22},\\
N_{1} &  =x_{11}\otimes W_{2},\text{ }N_{2}=x_{21}\otimes W_{2}.
\end{align}
Hence, $M_{1},M_{2},N_{1},N_{2}$ are included in the decomposable vector set
related to $\left(  W_{1},W_{2},\otimes\right)  .$ The construction is thus completed.

Before leaving this example, we would like to point out that $\left\{
\sigma_{x}^{A}\sigma_{x}^{B},\sigma_{z}^{A}\sigma_{z}^{B}\right\}  $ is also a
standard complete set of observables. In fact, we have%
\begin{align*}
\sigma_{x}^{A}\sigma_{x}^{B}\left\vert \psi^{\pm}\right\rangle _{AB} &
=\pm\left\vert \psi^{\pm}\right\rangle _{AB},\text{ }\sigma_{x}^{A}\sigma
_{x}^{B}\left\vert \phi^{\pm}\right\rangle _{AB}=\pm\left\vert \phi^{\pm
}\right\rangle _{AB},\\
\sigma_{z}^{A}\sigma_{z}^{B}\left\vert \psi^{\pm}\right\rangle _{AB} &
=-\left\vert \psi^{\pm}\right\rangle _{AB},\text{ }\sigma_{z}^{A}\sigma
_{z}^{B}\left\vert \phi^{\pm}\right\rangle _{AB}=+\left\vert \phi^{\pm
}\right\rangle _{AB}.
\end{align*}
It follows that $\left(  \sigma_{x}^{A}\sigma_{x}^{B},\sigma_{z}^{A}\sigma
_{z}^{B}\right)  $ is a standard complete set of observables with the same
characteristic sets $\left\{  M_{1},M_{2}\right\}  ,$ $\left\{  N_{1}%
,N_{2}\right\}  $ as defined above and the above constructed
$\left( A_{1},A_{2}\right)  $ is also an inner product compatible
TPP containing $\left\{
\sigma_{x}^{A}\sigma_{x}^{B},\sigma_{z}^{A}\sigma_{z}^{B}\right\}
.$ In this sense, we may well call $\left(
R_{x}(\pi),S_{z}^{2}\right)  $ and $\left(
\sigma_{x}^{A}\sigma_{x}^{B},\sigma_{z}^{A}\sigma_{z}^{B}\right)
$ equivalent.

\subsection{Entanglement in Bargmann Space}

Let $W$ be the space $%
\Bbb{C} \left[  x_{1},x_{2}\right]  $ of two variable polynomial
functions that span a
Berrgmann space of rank 2 \cite{Barg}. Notice that here an element of $%
\Bbb{C} \left[  x_{1},x_{2}\right]  $ is regarded as a function
rather than a polynomial in the indeterminates $x_{1}$ and
$x_{2}.$In studying this example, we have in mind the composite
system of two one dimensional subsystems. Indeed, $W$ can be
viewed in some way as a subspace of the space of states of such a
system, $x_{1}$ and $x_{2}$ understood as coordinates of the two
subsystems. This point of view was most recently casted on the
narrowing effects of wave packets of two free particles due to
their relative entanglement. 

Obviously, $\left\{  x_{1}^{j}x_{2}^{i}|j,i=0,1,\cdots\right\}  $ is a basis
of $W.$ Take $W_{1}=%
\Bbb{C}
\left[  x_{1}\right]  ,$ $W_{2}=%
\Bbb{C} \left[  x_{2}\right]  $ and define the bilinear map
$\otimes_{1}$from
$W_{1}\times W_{2}$ to $W$ : $x_{1}^{j}\otimes_{1}x_{2}^{i}=x_{1}^{j}x_{2}%
^{i}.$ Then $\left(  W_{1},W_{2},\otimes_{1}\right)  $ is a TPS of $W,$ and
actually this TPS is taken for granted. But we notice that if we define a
bilinear map $\otimes_{1}^{\prime}$: $W_{1}\times W_{2}\longrightarrow W$ such
that $x_{1}^{j}\otimes_{1}^{\prime}x_{2}^{i}=\alpha_{ji}x_{1}^{j}x_{2}^{i}$
where $\alpha_{ji}$ is a nonzero complex number, then $\left(  W_{1}%
,W_{2},\otimes_{1}^{\prime}\right)  $ is also a TPS. The state $x_{1}^{j}%
x_{2}^{i}$ is a product state with respect to both of the two tensor product
structures. Nevertheless, the decomposable vector states related to the two
tensor product structures are not identical when all $\alpha_{ji}$'s are not
identical. For example, if $\alpha_{11}=\alpha_{12}=\alpha_{21}=1,$ but
$\alpha_{22}=2,$ then
\begin{align}
x_{1}x_{2}+x_{1}x_{2}^{2}+x_{1}^{2}x_{2}+x_{1}^{2}x_{2}^{2} &  =\left(
x_{1}+x_{1}^{2}\right)  \otimes_{1}\left(  x_{2}+x_{2}^{2}\right)
,\nonumber\\
x_{1}x_{2}+x_{1}x_{2}^{2}+x_{1}^{2}x_{2}+x_{1}^{2}x_{2}^{2} &  =x_{1}%
\otimes_{1}^{\prime}\left(  x_{2}+x_{2}^{2}\right)  +x_{1}^{2}\otimes
_{1}^{\prime}\left(  x_{2}+\frac{1}{2}x_{2}^{2}\right)  .
\end{align}
So this is a product state with respect to $\left(  W_{1},W_{2},\otimes
_{1}\right)  $ but an entangled state with respect to $\left(  W_{1}%
,W_{2},\otimes_{1}^{\prime}\right)  .$ This result is no surprise. In fact,
when all $\alpha_{ji}$'s are not identical, generally speaking, $\left(
W_{1},W_{2},\otimes_{1}\right)  $ and $\left(  W_{1},W_{2},\otimes_{1}%
^{\prime}\right)  $ are associated with inequivalent tensor product
partitions. This being true, the result is then implied by\textit{\ }Lemma
7.1. This point can be argued as follows.

For simplicity, we suppose that $\alpha_{j0}=\alpha_{0i}=1$ for all $j,i.$ Let
$M_{i}$ be the space spanned by $\left\{  x_{1}^{j}x_{2}^{i}|j=0,1,\cdots
\right\}  ,$ $N_{j}$ the space spanned by $\left\{  x_{1}^{j}x_{2}%
^{i}|i=0,1,\cdots\right\}  .$ For $\left\{  \alpha_{ji}\right\}  \subseteq%
\Bbb{C} ,$ define a TPP $\left(  A_{1}\left(  \alpha\right)
,A_{2}\left( \alpha\right)  \right)  $ such that (1) $\left\{
M_{i}\right\}  ,\left\{ N_{j}\right\}  $ are irreducible component
sets for $A_{1}\left( \alpha\right)  $ and $A_{2}\left(
\alpha\right)  $ respectively; (2) $\left\{
\alpha_{ji}x_{1}^{j}x_{2}^{i}\right\}  $ is a standard basis
associated with $\left\{  M_{i}\right\}  ,\left\{  N_{j}\right\}
.$ Now one can check that $\left(
W_{1},W_{2},\otimes_{1}^{\prime}\right)  $ is associated with
$\left(  A_{1}\left(  \alpha\right)  ,A_{2}\left(
\alpha\right)  \right)  $ and%
\[
A_{1}\left(  \alpha\right)  =\varphi\cdot A_{1}\left(  1\right)  \cdot
\varphi^{-1},\text{ }A_{2}\left(  \alpha\right)  =\varphi\cdot A_{2}\left(
1\right)  \cdot\varphi^{-1}%
\]
where $\varphi$ is an automorphism of $W$ such that
\[
\varphi\left(  x_{1}^{j}x_{2}^{i}\right)  =\alpha_{ji}x_{1}^{j}x_{2}^{i}.
\]
When all $\alpha_{ji}$'s are not identical, $\varphi$ is not an identity map.
So it is highly possible that $\left(  A_{1}\left(  \alpha\right)
,A_{2}\left(  \alpha\right)  \right)  $ and $\left(  A_{1}\left(  1\right)
,A_{2}\left(  1\right)  \right)  $ are not equivalent.

\subsection{Entanglement with respect to Mass of Center Coordinate}

Now let us study another TPS of $W$ as a special case of the above example.
The discussion is motivated by the simple consideration that a separable wave
function of bi-particle system with respect to the two position coordinates
can be regarded as an entangling one with respect to the center of mass and
relative coordinates.

Let $X$ be the ``coordinate of center of mass", $x$ the ``relative
coordinate":
\begin{equation}
X=\frac{1}{2}(x_{1}+x_{2}),x=x_{1}-x_{2}.
\end{equation}
One can check that $\left\{  X^{j}x^{i}|j,i=0,1,\cdots\right\}  $ is also a
basis of $W.$ Consider the operators $X\cdot\frac{\partial}{\partial X}$ and
$x\cdot\frac{\partial}{\partial x},$ where $\frac{\partial}{\partial X}$ and
$\frac{\partial}{\partial x}$ are derivatives with respect to $X$ and $x$
respectively . By definition we have
\begin{align}
\left(  X\cdot\frac{\partial}{\partial X}\right)  X^{j}x^{i}  &  =jX^{j}%
x^{i},\\
\text{ }\left(  x\cdot\frac{\partial}{\partial x}\right)  X^{j}x^{i}  &
=iX^{j}x^{i}.\nonumber
\end{align}
Let $M_{i},N_{j}$ be the space spanned by $\left\{  X^{j}x^{i}|j=0,1,\cdots
\right\}  $ and $\left\{  X^{j}x^{i}|i=0,1,\cdots\right\}  $ respectively.
Then it follows that $\left(  X\cdot\frac{\partial}{\partial X},x\cdot
\frac{\partial}{\partial x}\right)  $ is a standard complete set of operators,
and $\left\{  M_{i}\right\}  ,$ $\left\{  N_{j}\right\}  $ are the
characteristic sets of $X\cdot\frac{\partial}{\partial X}$ and $x\cdot
\frac{\partial}{\partial x}$ respectively.

According to \textit{Proposition 4.1}, there is a TPP containing $\left(
X\cdot\frac{\partial}{\partial X},x\cdot\frac{\partial}{\partial x}\right)  .$
Let us explicitly construct such a TPP. We define the extended subalgebras
$A_{1},A_{2}\subseteq End(W)$ as follows. An element $a$ belongs to $A_{1}$ if
and only if
\begin{equation}
aX^{j}x^{i}=\sum_{k}X^{k}x^{i}a_{kj},
\end{equation}
and an element $b$ belongs to $A_{2}$ if and only if
\begin{equation}
bX^{j}x^{i}=\sum_{k}X^{j}x^{k}b_{ki},
\end{equation}
where $a_{kj}\in%
\Bbb{C}
$ is independent of $i,$ $b_{ki}\in%
\Bbb{C} $ is independent of $j,$ and $\left\{
k|a_{kj}\neq0\right\}  ,$ $\left\{ k|b_{ki}\neq0\right\}  $ are
both finite sets for each $j$ and each $i$
respectively. According to the proof of Proposition 4.1, $\left(  A_{1}%
,A_{2}\right)  $ is a TPP containing $\left(  X\cdot\frac{\partial}{\partial
X},x\cdot\frac{\partial}{\partial x}\right)  .$

Now take $W_{1}^{\prime}$ and $W_{2}^{\prime}$ to be the subspaces spanned by
$\left\{  X^{j}\right\}  $ and $\left\{  x^{i}\right\}  $ respectively, and
define a bilinear map $\otimes_{2}:$ $W_{1}^{\prime}\times W_{2}^{\prime
}\longrightarrow W$ such that $X^{j}\otimes_{2}x^{i}=X^{j}x^{i}.$ It is then
readily check that $\left(  W_{1}^{\prime},W_{2}^{\prime},\otimes_{2}\right)
$ is a TPS associated with $\left(  A_{1},A_{2}\right)  .$

Finally we point out that the decomposable vector set related to $\left(
W_{1}^{\prime},W_{2}^{\prime},\otimes_{2}\right)  $ and that related to
$\left(  W_{1},W_{2},\otimes_{1}\right)  $ are different. For example, we
consider the state $x_{1}x_{2}.$ We have%
\begin{align}
x_{1}x_{2} &  =\frac{2X+x}{2}\frac{2X-x}{2}\nonumber\\
&  =X^{2}\otimes_{2}1-\frac{1}{4}\left(  1\otimes_{2}x^{2}\right)  .
\end{align}
So it is an entangled state with respect to $\left(  W_{1}^{\prime}%
,W_{2}^{\prime},\otimes_{2}\right)  .$ But it is a product state with respect
to $\left(  W_{1},W_{2},\otimes_{1}\right)  .$

The above argument demonstrate a simple but profound physical fact
about the relativity of entanglement : generally speaking, the
factorized wave function
$\Psi(x_{1},x_{2})=\Psi_{1}(x_{1})\Psi_{2}(x_{2})$ is entangled
with respect to the ''coordinate of center of mass" $\left(
X,x\right)  $ since generally
there are no functions $\Phi_{1},\Phi_{2}$ such that%

\begin{equation}
\Psi(x_{1},x_{2})=\Phi_{1}(X)\Phi_{2}(x)
\end{equation}
though we do have%
\[
\Psi(x_{1},x_{2})=\Psi_{1}(X+\frac{x}{2})\Psi_{2}(X-\frac{x}{2}).
\]

\textit{Remark} If we interprets $X,x$\ as creation operators and
$\frac{\partial}{\partial X},\frac{\partial}{\partial x}$\ as
annihilation operators respectively, then the above discussion is
applicable to settling down the issue of entanglement of Fock
states, as promised in the introduction.

\section{ Concluding Remarks}

We have presented a rigorous algebraic description for the relativity of
quantum entanglement due to the non-uniqueness of TPS of a vector space.
Physically, there are many ways to subdivide the Hilbert space of a large
system according to various physical purposes. In practice, different
partitions correspond to different choices of observables in the measurement.
According to the above discussion, this means that the notion of entanglement
depends on the definition of tensor product in association with the subsystem
partition. This reveals the seemingly exotic fact that multi-particle states
that are entangled with respect to some subsystem partition may be separable
with respect to different observations in the measurement. For example, a
symmetrized state for two boson system is obviously an entangled state in the
coordinate representation, but it is a tensor product of two number state with
respect to some TPS. We think that it is safe to say that the present paper
has made clear the cloudy physical concept-- quantum entanglement in a
mathematical way.

Finally, we would like to remark that in this paper we have made
efforts not to leave out the infinite dimensional case, avoiding
the argument's being too restrictive and excluding many physically
interesting examples. But on the other hand, we have completely
sacrificed topology for mathematical simplicity. So from
mathematical point, especially analytical point of view, the
present paper has left much to be desired. It seems desirable to
study the inner product compatible TPS of an infinite dimensional
Hilbert space $H $. Then we will have to consider the topology of
$L(H),$ the set of linear operators on $H,$ and to study some kind
of partition of $L(H)$ related to the inner product compatible TPS
of $H$ we might need to enter the field of operator algebra.

\textit{We thank Paolo Zanardi for stimulating discussions. This
work is supported by the NSFC and the knowledge Innovation Program
(KIP) of the Chinese Academy of Sciences. It is also founded by
the National Fundamental Research Program of China with No.
001GB309310. }

\end{document}